\documentclass[12pt]{article}

\usepackage{graphicx}
\usepackage{graphics}
\usepackage{amssymb}
\usepackage{amsmath}

\usepackage{epsf,amsfonts,hyperref}

\renewcommand{\appendix}[1]{
    \addtocounter{section}{1}
    \setcounter{equation}{0}
    \renewcommand{\thesection}{\Alph{section}}
    \section*{Appendix \thesection\protect\indent #1}
    \addcontentsline{toc}{section}{Appendix \thesection\ \ \ #1}
}
\newcommand\encadremath[1]{\vbox{\hrule\hbox{\vrule\kern8pt 
\vbox{\kern8pt \hbox{$\displaystyle #1$}\kern8pt} 
\kern8pt\vrule}\hrule}}
\def\enca#1{\vbox{\hrule\hbox{
\vrule\kern8pt\vbox{\kern8pt \hbox{$\displaystyle #1$}
\kern8pt} \kern8pt\vrule}\hrule}}

\newcommand\figureframex[3]{
\begin{figure}[bth]
\hrule\hbox{\vrule\kern8pt 
\vbox{\kern8pt \vbox{
\begin{center}
{\mbox{\epsfxsize=#1.truecm\epsfbox{#2}}}
\end{center}
\caption{#3}
}\kern8pt} 
\kern8pt\vrule}\hrule
\end{figure}
}
\newcommand\figureframey[3]{
\begin{figure}[bth]
\hrule\hbox{\vrule\kern8pt 
\vbox{\kern8pt \vbox{
\begin{center}
{\mbox{\epsfysize=#1.truecm\epsfbox{#2}}}
\end{center}
\caption{#3}
}\kern8pt} 
\kern8pt\vrule}\hrule
\end{figure}
}

\makeatletter
\@addtoreset{equation}{section}
\makeatother
\newtheorem{theorem}{Theorem}[section]

\newtheorem{remark}{Remark}[section]
\newtheorem{proposition}{Proposition}[section]
\newtheorem{lemma}{Lemma}[section]
\newtheorem{corollary}{Corollary}[section]
\newtheorem{definition}{Definition}[section]
\def\br{\begin{remark}\rm\small}
\def\er{\end{remark}}
\def\bt{\begin{theorem}}
\def\et{\end{theorem}}
\def\bd{\begin{definition}}
\def\ed{\end{definition}}
\def\bp{\begin{proposition}}
\def\ep{\end{proposition}}
\def\bl{\begin{lemma}}
\def\el{\end{lemma}}
\def\bc{\begin{corollary}}
\def\ec{\end{corollary}}
\def\beaq{\begin{eqnarray}}
\def\eeaq{\end{eqnarray}}
\newcommand{\proof}[1]{{\noindent \bf proof:}\par
{#1} $\square$}

\newcommand{\beq}{\begin{equation}}
\newcommand{\eeq}{\end{equation}}
\newcommand{\bea}{\begin{eqnarray}}
\newcommand{\eea}{\end{eqnarray}}

\renewcommand{\and}{{\qquad {\rm and} \qquad}}

\newcommand{\virg}{{\qquad , \qquad}}

\newcommand{\Res}{\mathop{\,\rm Res\,}}

\newcommand{\om}{\omega}

\newcommand{\ee}[1]{{{\rm e}^{#1}}}

\renewcommand{\d}{{{\partial}}}

\newcommand{\Pint}{{\int\kern -1.em -\kern-.25em}}

\renewcommand{\L}{\Lambda}

\newcommand{\ovl}{\overline}
\newcommand{\genus}{{\mathfrak g}}

\newcommand{\acycle}{{\cal A}}
\newcommand{\bcycle}{{\cal B}}

\newcommand\spcurve{{\cal S}}
\newcommand\curve{{\cal C}}

\newcommand\Arond{{\stackrel{\circ}{A}}}
\newcommand\Brond{{\stackrel{\circ}{B}}}

\newcommand\Erond{{\stackrel{\circ}{E}}}

\newcommand\wrond{{\stackrel{\circ}{w}}}

\newcommand\CL{{{\hat \Lambda}}}
\newcommand\modsp{{\cal M}}

\newcommand\Ber{{{\cal B}}}

\newcommand\bpt{{\mathfrak b}}

\newcommand\CYX{{\mathfrak X}}

\begin{document}
\sloppy


\pagestyle{empty}

\hfill IPHT-T11/200
\addtolength{\baselineskip}{0.20\baselineskip}
\begin{center}
\vspace{26pt}
{\large \bf {Invariants of spectral curves and intersection theory of moduli spaces of complex curves}}
\newline
\vspace{26pt}

{\sl B.\ Eynard}\hspace*{0.05cm}\footnote{ E-mail: bertrand.eynard@cea.fr }\\
\vspace{6pt}
Institut de Physique Th\'{e}orique, CE Saclay,\\
F-91191 Gif-sur-Yvette Cedex, France.\\
D\'{e}partement de Math\'{e}matiques, Universit\'{e} of Gen\`{e}ve \\
2-4, rue du Li\`{e}vre, Case postale 64, 1211 Gen\`{e}ve 4, Switzerland.\\
\end{center}

\vspace{20pt}
\begin{center}
{\bf Abstract}

To any spectral curve $\spcurve$, we associate a topological class $\hat\Lambda(\spcurve)$ in a moduli space $\modsp_{g,n}^\bpt$ of "$\bpt$--colored" stable Riemann surfaces of given topology (genus $g$, $n$ boundaries), whose integral coincides with the topological recursion invariants $W_{g,n}(\spcurve)$ of the spectral curve $\spcurve$.
This formula can be viewed as a generalization of the ELSV formula (whose spectral curve is the Lambert function and the associated class is the Hodge class), or Mari\~no--Vafa formula (whose spectral curve is the mirror curve of the framed vertex, and the associated class is the product of 3 Hodge classes), but for an arbitrary spectral curve.
In other words, to a B-model (i.e. a spectral curve) we systematically associate a mirror A-model (integral  in a moduli space of "colored" Riemann surfaces). We find that the mirror map, i.e. the relationship between the A-model moduli and B-model moduli, is realized by the Laplace transform.


\end{center}
%





\vspace{26pt}
\pagestyle{plain}
\setcounter{page}{1}


\section{Introduction}

In the past few years, many developments have unearthed a deep and fascinating relationship between integrable systems, algebraic geometry, combinatorics, enumerative geometry and random matrices, and much is yet to be understood.

In particular, an integrable system can be encoded by its "spectral curve" (i.e. the locus of eigenvalues of a Lax operator), which is a plane analytical curve embedded in $\mathbb C\times \mathbb C$, for example given by its equation, like $y=\sin{(\sqrt x)}$, or $\ee{x} = y\,\ee{-y}$.

In \cite{EOFg} (led by recent developments in random matrix theory), it was proposed to define a sequence of "invariants" $W_n^{(g)},\, g=0,1,2,\dots, \,\, n=0,1,2,\dots$ associated to a spectral curve.
For many specific examples of spectral curves, those invariants had an enumerative geometry interpretation, as "counting" surfaces in a moduli space of Riemann surfaces.

For example: 

- for the spectral curve  $\mathbf{ y=\sqrt x}$, the $W_n^{(g)}$'s are the generating functions of Witten--Kontsevich intersection numbers on the moduli space $\modsp_{g,n}$ of Riemann surfaces of genus $g$ with $n$ marked points \cite{Kontsevich1992}. Writing $z_i=\sqrt{x_i}$ we have
$$
W_n^{(g)}
= (-2)^{2g-2+n}\sum_{d_1+\dots+d_n=3g-3+n}\,\left<\prod_{i=1}^n \tau_{d_i}\right>_{\modsp_{g,n}}\,\prod_{i=1}^n \frac{(2d_i+1)!!\,\,dz_i}{z_i^{2d_i+2}} \,.
$$

- for the spectral curve $\mathbf{y=\frac{1}{2\pi}\sin{(2\pi\sqrt x})}$, the $W_n^{(g)}$'s are the Laplace transforms of Weil-Petersson volumes of the moduli spaces $\modsp_{g,n}$ (see \cite{Mulase2006, Eynard2007}). Writing $z_i=\sqrt{x_i}$ we have
\bea\nonumber
W_n^{(g)}
&=& \prod_{i=1}^n \int_0^\infty L_idL_i\,\ee{-z_i L_i}\,\,{\rm Vol}(\modsp_{g,n}(L_1,\dots,L_n)) \cr
&=& (-2)^{2g-2+n}\sum_{d_0+d_1+\dots+d_n=3g-3+n}\,\frac{1}{d_0!}\,\left<(2\pi^2\kappa_1)^{d_0}\,\prod_{i=1}^n \tau_{d_i}\right>_{\modsp_{g,n}}\,\prod_{i=1}^n \frac{(2d_i+1)!!\,\,dz_i}{z_i^{2d_i+2}} \, \cr
&=& (-2)^{2g-2+n}\sum_{d_1,\dots,d_n}\,\left<\ee{2\pi^2\kappa_1}\,\prod_{i=1}^n \tau_{d_i}\right>_{\modsp_{g,n}}\,\prod_{i=1}^n \frac{(2d_i+1)!!\,\,dz_i}{z_i^{2d_i+2}} \,. \cr
\eea

- for the spectral curve $\mathbf{\ee{x}=y\,\ee{-y}}$, it was proposed by Bouchard and Mari\~no \cite{BM} that the $W_n^{(g)}$'s are the generating functions of simple Hurwitz numbers ${\cal H}_{g,\mu}$, which was then proved in \cite{Borot1,Eynardb}. And simple Hurwitz numbers can themselves be translated into integrals of the Hodge class $\Lambda(1)$ in the moduli space of curves, through ELSV formula \cite{EKEDAHL1999}:
\bea\nonumber
W_n^{(g)} 
&=& (-1)^n\,\sum_{\mu\,, \, \ell(\mu)=n}\,{\cal H}_{g,\mu}\,\prod_{i=1}^n \mu_i\,\ee{-\mu_i\,x_i}\,dx_i  \cr
&=&  (-1)^n\,\sum_{\mu_1,\dots,\mu_n}\,\left< \Lambda(1)\,\prod_{i=1}^n \frac{\mu_i}{1-\mu_i\psi_i}\right>_{\modsp_{g,n}}\,\prod_{i=1}^n \frac{\mu_i^{\mu_i}}{\mu_i !}\,\ee{-\mu_i\,x_i}\,dx_i  \, .\cr
\eea

- More generally, it was initially proposed by Mari\~no \cite{Mar1} and refined by Bouchard--Klemm--Mari\~no--Pasquetti (BKMP) \cite{BKMP}, that if we choose the spectral curve to be the mirror curve of a toric Calabi--Yau 3--fold $\mathfrak X$, then the $W_n^{(g)}$'s should be generating functions for the Gromov--Witten invariants of $\mathfrak X$, i.e. they enumerate maps from a Riemann surface of genus $g$ into $\mathfrak X$  with $n$ boundaries into a Lagrangian submanifold of $\mathfrak X$. This BKMP conjecture was proved to low genus for many spaces $\mathfrak X$, and to all genus only  for the case $\mathfrak X=\mathbb C^3$ in \cite{ChenLin2009, ZhouJian2009}.

\bigskip

Then, if a spectral curve is not among the list of "known examples" (the list above is not exhaustive, there are also more known examples associated to matrix models \cite{EOFg}, combinatorics of maps and combinatorics of 2D and 3D partitions, counting Grothendieck's dessins d'enfants \cite{Mulase2010}, ...), the natural question is:

\smallskip
{\bf - do the $W_n^{(g)}$'s of an arbitrary spectral curve have a meaning in terms of enumerative geometry, counting the  "volume of some moduli space of surfaces" ?}

\medskip
In \cite{Eynard2011} we proposed a partial answer for all spectral curves having only one branchpoint.
The goal of the present article is to extend that to an arbitrary number of branchpoints.
We shall thus define a compact moduli space $\ovl\modsp^\bpt_{g,n}$ of "colored" Riemann surfaces of genus $g$ with $n$ marked points, and some (cohomology class of) differential forms $\CL(\spcurve)$ and $\hat B(z,1/\psi)$ on it, so that $W_n^{(g)}$ is indeed an integral on that moduli--space.

We shall prove in theorem \ref{mainth} that:
\beq
\encadremath{
W_n^{(g)}(\spcurve;z_1,\dots,z_n) = 2^{{\rm dim}\,\modsp^\bpt_{g,n}}\int_{\ovl\modsp^\bpt_{g,n}}\,\CL(\spcurve)\,\prod_{i=1}^n\hat B(z_i,1/\psi_i)
}\eeq
where notations are explained below.

We shall see that the term $\hat B(z_i,1/\psi_i)$ or more precisely its Laplace transform $\int\ee{-\mu_i x(z_i)}\hat B(z_i,1/\psi_i)$ is the analogous of $\frac{\mu_i}{1+\mu_i\psi_i}$ in the ELSV formula, and that it indeed reduces to it if we choose the spectral curve to be the Lambert function.
In some sense our formula generalizes the ELSV formula.

\section{Definition and notations: spectral curves and their invariants}

The goal of this article is to show that invariants of an arbitrary spectral curve can be written in terms of intersection numbers, so we first recall the definition of symplectic invariants and their descendants.

\bd[Spectral curve]
a spectral curve $\spcurve=(\curve,x,y,B)$, is the data of:

$\bullet$ a Riemann surface $\curve$ (not necessarily compact neither connected),

$\bullet$ two analytical function $x:\curve\to\mathbb C$, $y:\curve\to\mathbb C$,

$\bullet$ a Bergman kernel $B$, i.e. a symmetric 2nd kind bilinear meromorphic differential, having a double pole on the diagonal and no other pole, and normalized (in any local coordinate $z$) as:
\beq
B(z_1,z_2) \mathop{{\sim}}_{z_2\to z_1} \frac{dz_1\otimes dz_2}{(z_1-z_2)^2} + {\rm analytical}.
\eeq

\bigskip

Moreover, the spectral curve $\spcurve$ is called regular if the meromorphic form $dx$ has a finite number of zeroes on $\curve$, and they are simple zeroes, and $dy$ doesn't vanish at the zeroes of $dx$. In other words, locally near a branchpoint $a$, $y$ behaves like a square root of $x$:
\beq\label{bpysqrtx}
y(z) \mathop{{\sim}}_{z\to a} y(a) + y'(a)\,\sqrt{x(z)-x(a)} + O(x(z)-x(a)) \qquad , \,\, y'(a)\neq 0.
\eeq

\ed

From now on, all spectral curves considered shall always be chosen to be regular\footnote{A generalized definition of symplectic invariants for non--regular spectral curves was also introduced in \cite{PratsFerrer2010}, but for simplicity, we consider only regular spectral curves here.}.

\bigskip

In \cite{EOFg}, it was proposed how to associate to a regular spectral curve, an infinite sequence of symmetric meromorphic $n$-forms, and a sequence of complex numbers $F_g(\spcurve)$.
The definition is given by a recursion, often called "topological recursion", which we recall:

\bd[Invariants $W_n^{(g)}(\spcurve)$]
Let $\spcurve=(\curve,x,y,B)$ be a regular spectral curve.
Let $a_1,\dots,a_\bpt$ be its branchpoints (zeroes of $dx$ in $\curve$).
We define
\beq
W_1^{(0)}(\spcurve;z) = y(z)\,dx(z),
\eeq
\beq
W_2^{(0)}(\spcurve;z_1,z_2) = B(z_1,z_2),
\eeq
and for $2g-2+(n+1)>0$:
\bea
W_{n+1}^{(g)}(\spcurve;z_1,\dots,z_n,z_{n+1})
&=& \sum_{i=1}^\bpt \Res_{z\to a_i} K(z_{n+1},z)\,\Big[ W_{n+2}^{(g-1)}(z,\bar z,z_1,\dots,z_n) \cr
&& \qquad + \sum_{h=0}^g\sum'_{I\uplus J = \{z_1,\dots,z_n\}} W_{1+\#I}^{(h)}(z,I)\,W_{1+\#J}^{(g-h)}(z,J) \Big] \cr
\eea
where the prime in $\sum_h\sum_{I\uplus J}'$ means that we exclude from the sum the terms $(h=0,I=\emptyset)$ and $(h=g,J=\emptyset)$, 
and
where $\bar z$ means the other branch of the square-root in \eqref{bpysqrtx} near a branchpoint $a_i$, i.e. if $z$ is in the vicinity of $a_i$, $\bar z\neq z$ is the other point in the vicinity of $a_i$ such that
\beq
x(\bar z)=x(z),
\eeq
and thus $y(\bar z) \sim y(a) - y'(a)\sqrt{x(z)-x(a)}$.
The recursion kernel $K(z_{n+1},z)$ is defined as
\beq
K(z_{n+1},z) = \frac{\int_{z'=\bar z}^z B(z_{n+1},z')}{2(y(z)-y(\bar z))\,dx(z)}
\eeq
$K$ is a 1-form in $z_{n+1}$ defined on $\curve$ with a simple pole at $z_{n+1}=z$ and at $z_{n+1}=\bar z$, and in $z$ it is the inverse of a 1-form, defined only locally near branchpoints, and it has a simple pole at $z=a_i$.

\medskip

We also define for $g\geq 2$:
\beq
F_g(\spcurve)=W_0^{(g)}(\spcurve) = \frac{1}{2-2g}\,\sum_{i=1}^\bpt \Res_{z\to a_i}\,\,W_1^{(g)}(\spcurve;z)\,\,\left(\int_{z'=a_i}^z\,y(z')dx(z')\right).
\eeq

\ed

With this definition, $F_g(\spcurve)\in\mathbb C$ is a complex number associated to $\spcurve$, sometimes called the $g^{\rm th}$ symplectic invariant of $\spcurve$, and $W_n^{(g)}(\spcurve;z_1,\dots,z_n)$ is a symmetric multilinear differential $\in T^*(\curve)\otimes\dots\otimes T^*(\curve)$, sometimes called the $n^{\rm th}$ descendant of $F_g$. Very often we denote $F_g=W_0^{(g)}$.
If $2-2g-n<0$, $W_n^{(g)}$ is called stable, and otherwise unstable, the only unstable cases are $F_0, F_1, W_1^{(0)}, W_2^{(0)}$.
 For $2-2g-n<0$, $W_n^{(g)}$ has poles only at branchpoints (when some $z_k$ tends to a branchpoint $a_i$), without residues, and the degrees of the poles are $\leq 6g+2n-4$.

\smallskip

It is also possible to define $F_0$ and $F_1$, see \cite{EOFg}, but we shall not use them here.

\medskip

Those invariants $F_g$ and $W_n^{(g)}$'s have many fascinating properties, in particular related to integrability, to modular functions, and to special geometry, and we refer the reader to \cite{EOFg, Eynard2008}.


\section{Intersection numbers}

Our goal now is to relate those $W_n^{(g)}$'s to intersection numbers in moduli spaces of curves, so let us first introduce basic concepts.

\subsection{Definitions}

\begin{figure}[t]
\centering
\includegraphics[height=5cm]{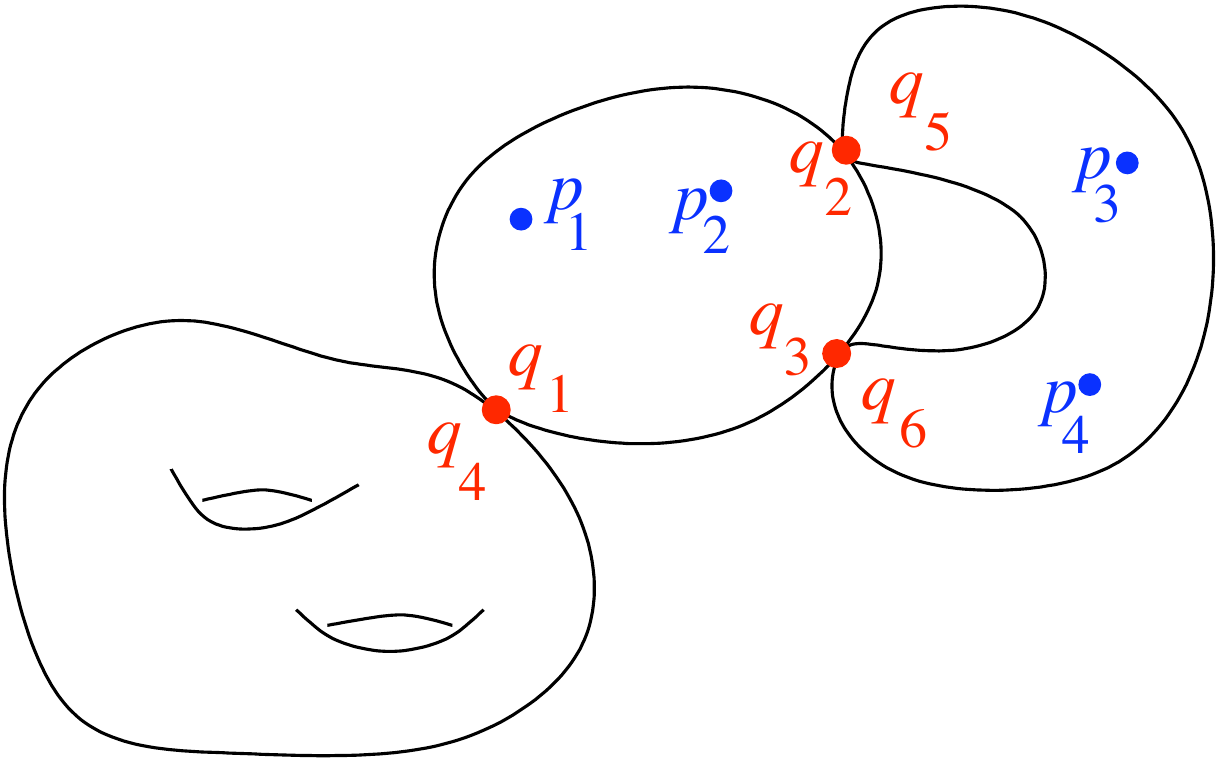} 
\caption{A stable curve in $\ovl\modsp_{g,n}$ can be smooth or nodal. Here we have an example in $\ovl\modsp_{3,4}$ of a stable curve of genus $g=3$, with $n=4$ marked points $p_1,\dots, p_4$, and made of 3 components, glued by 3 nodal points. Each nodal point is a pair of marked points $(q_i,q_j)$. Each component is a smooth Riemann surface of some genus $g_i$, and with $n_i$ marked or nodal points. Stability means that for each component $\chi_i=2-2g_i-n_i<0$. Here, one component has genus 2 and 1 nodal point $q_4$ so $\chi=-3$, another component is a sphere with 2 marked points $p_1,p_2$ and 3 nodal points $q_1,q_2,q_3$ i.e. $\chi=-3$, and the last component is a sphere with 2 marked points $p_3,p_4$ and 2 nodal points $q_5,q_6$ so $\chi=-2$. The total Euler characteristics is $\chi=-3-3-2=-8$ which indeed corresponds to $2-2g-n$ for a Riemann surface of genus $g=3$ with $n=4$ marked points.\label{fignodalmap1}
} 
\end{figure}

Let ${\cal M}_{g,n}$ be the moduli space of complex curves of genus $g$ with $n$ marked points.
It is a complex orbifold (manifold quotiented by a group of symmetries), of dimension
\beq
{\rm dim}\,{\cal M}_{g,n}=d_{g,n}=3g-3+n.
\eeq
Each element $(\Sigma,p_1,\dots,p_n)\in {\cal M}_{g,n}$ is a smooth complex curve $\Sigma$ of genus $g$ with $n$ smooth distinct marked points $p_1,\dots,p_n$.
$\modsp_{g,n}$ is not compact because the limit of a family of smooth curves may be non--smooth, some cycles may shrink, or some marked points may collapse in the limit.
The Deligne--Mumford compactification $\overline\modsp_{g,n}$ of $\modsp_{g,n}$ also contains stable nodal curves of genus $g$ with $n$ marked points (a nodal curve is a set of smooth curves glued at nodal points, and thus nodal points are equivalent to pairs of marked points, and stability means that each punctured component curve has an Euler characteristics $<0$), see fig \ref{fignodalmap1}.
$\overline\modsp_{g,n}$ is then a compact space.

Let ${\cal L}_i$ be the cotangent bundle at the marked point $p_i$, i.e. the bundle over $\overline{\cal M}_{g,n}$ whose fiber is the cotangent space $T^*(p_i)$ of $\Sigma$ at $p_i$.
It is customary to denote its first Chern class:
\beq
\psi_i=\psi(p_i)=c_1({\cal L}_i).
\eeq
$\psi_i$ is (the cohomology equivalence class modulo exact forms, of) a 2-form on $\ovl{\cal M}_{g,n}$.
Since $\dim_{\mathbb R} \ovl\modsp_{g,n}= 2 \dim_{\mathbb C} \ovl\modsp_{g,n}= 6g-6+2n$, it makes sense to compute the integral of the exterior product of $3g-3+n\,$ 2-forms, i.e. to compute the "intersection number"
\beq
\left<\psi_1^{d_1}\dots \psi_n^{d_n}\right>_{g,n} = \int_{[\overline{\cal M}_{g,n}]^{\rm vir}}\psi_1^{d_1}\dots \psi_n^{d_n}
\eeq
on the Deligne--Mumford compactification $\overline{\cal M}_{g,n}$ of ${\cal M}_{g,n}$ (or more precisely, on a virtual cycle $[]^{\rm vir}$ of $\overline{\cal M}_{g,n}$, taking carefully account of the non-smooth curves at the boundary of ${\cal M}_{g,n}$), provided that
\beq
\sum_i d_i=d_{g,n}=3g-3+n.
\eeq
If this equality is not satisfied we define $\left<\psi_1^{d_1}\dots \psi_n^{d_n}\right>_{g,n} =0$.

\medskip

More interesting characteristic classes and intersection numbers are defined as follows.
Let (we follow the notations of \cite{Liu2009}, and refer the reader to it for details)
$$
\pi:\overline{\cal M}_{g,n+1}\to \overline{\cal M}_{g,n}
$$
be the forgetful morphism (which forgets the last marked point), and let $\sigma_1,\dots,\sigma_n$ 
be the canonical sections of $\pi$, and $D_1,\dots,D_n$ be the corresponding divisors in $\overline{\cal M}_{g,n+1}$. Let $\om_\pi$ be the relative dualizing sheaf.
We consider the following tautological classes on $\overline{\cal M}_{g,n}$:

$\bullet$ The $\psi_i$ classes (which are 2-forms), already introduced above:
$$ \psi_i = c_1(\sigma_i^*(\om_\pi)) $$
It is customary to use Witten's notation:
\beq\label{deftaud}
\psi_i^{d_i}=\tau_{d_i}.
\eeq

$\bullet$ The Mumford $\kappa_k$ classes \cite{Mumford1983, Arbarello1996}:
$$ \kappa_k = \pi_*(c_1(\om_\pi(\sum_i D_i))^{k+1} ) .$$
$\kappa_k$ is a $2k$--form. 
$\kappa_0$ is the Euler class, and in $\ovl{\cal M}_{g,n}$, we have
$$
\kappa_0=-\chi_{g,n}=2g-2+n.
$$
$\kappa_1$ is known as the Weil-Petersson form since it is given by $2\pi^2\kappa_1=\sum_i dl_i\wedge d\theta_i$ in the Fenchel-Nielsen coordinates $(l_i,\theta_i)$ in Teichm\"uller space \cite{Wolpert1983}.

In some sense, $\kappa$ classes are the remnants of the $\psi$ classes of (clusters of) forgotten points.
There is the formula \cite{Arbarello1996}:
\beq
\pi_* \psi_1^{d_1}\dots\psi_n^{d_n}\,\psi_{n+1}^{k+1}\, = \psi_1^{d_1}\dots\psi_n^{d_n}\,\kappa_k
\eeq
\beq
\pi_*\pi_* \psi_1^{d_1}\dots\psi_n^{d_n}\,\psi_{n+1}^{k+1}\,\psi_{n+2}^{k'+1}\, = \psi_1^{d_1}\dots\psi_n^{d_n}\,(\kappa_k\,\kappa_{k'}+\kappa_{k+k'})
\eeq
and so on...

\medskip
$\bullet$ The Hodge class  $\Lambda(\alpha)=1+\sum_{k=1}^g \,(-1)^k\,\alpha^{-k}\,c_k(\mathbb E)$ where $c_k(\mathbb E)$ is the $k^{\rm th}$ Chern class of the Hodge bundle $\mathbb E=\pi_*(\om_\pi)$.
Mumford's formula \cite{Mumford1983, FaberC.1998} says that
\beq\label{eqmumfordhodge}
\Lambda_{\rm Hodge}(\alpha)= \ee{\sum_{k\geq 1} { \frac{B_{2k}\,\alpha^{1-2k}}{2k(2k-1)}\,\,\left(\kappa_{2k-1}-\sum_i \psi_i^{2k-1}+\frac{1}{2}\sum_\delta \sum_j (-1)^j\,\,l_{\delta*} \psi^j\,\psi'^{2k-2-j}\right)}}
\eeq
where $\Ber_{k}$ is the $k^{\rm th}$ Bernoulli number, $\delta$ a boundary divisor (i.e. a cycle which can be pinched so that the pinched curve is a stable nodal curve, i.e. replacing the pinched cycle by a pair of marked points, all components have a strictly negative Euler characteristics), and $l_{\delta*}$ is the natural inclusion into the moduli spaces of each connected component. In other words $\sum_\delta l_{\delta*}$ adds a nodal point in all possible stable ways. 

\bigskip
In fact, all tautological classes in $\overline{\cal M}_{g,n}$ can be expressed in terms of $\psi$-classes or their pull back or push forward from some $\overline{\cal M}_{h,m}$ \cite{Bertram2006}.
Faber's conjecture \cite{FaberC.1998} (partly proved in \cite{Mulase2006} and \cite{Liu2009}) proposes an efficient method to compute intersection numbers of $\psi, \kappa$ and Hodge classes.

\subsection{Reminder 1 branch-point}

If $\spcurve_a=(\curve_a,x,y,B)$ is a spectral curve with only one branchpoint $a$, the following theorem was proved in \cite{Eynard2011}:

\bt[1 branchpoint]\label{th1bp}
\bea\label{Wng1bp}
W_n^{(g)}(\spcurve_a;z_1,\dots,z_n) 
&=& 2^{d_{g,n}}\,\sum_{d_1,\dots,d_n} \prod_{i=1}^n d\xi_{a,d_i}(z_i)\,\left< \psi_1^{d_1}\dots \psi_n^{d_n}\,\,\CL(\spcurve_a) \right>_{\modsp_{g,n}} \cr
&=& 2^{d_{g,n}}\,\left< \prod_{i=1}^n\, \hat B_{a}(z;1/\psi_i)\,\,\CL(\spcurve_a) \right>_{\modsp_{g,n}} \cr
\eea
or alternatively after Laplace transform:
\bea\label{Wng1bpLaplace}
&& \int_{\gamma_a^n}\,\, W_n^{(g)}(\spcurve_a;z_1,\dots,z_n) \prod_{i=1}^n\ee{-\mu_i\,x(z_i)}  \cr
&=& 2^{d_{g,n}}\,\prod_{i=1}^n \frac{2\sqrt{\pi}\,\ee{-\mu_ix(a)}}{\sqrt{\mu_i}}\,\, \left< \prod_{i=1}^n\, \left({\mu_i\over 1+\mu_i\psi_i}-\check B_{a,a}(\mu_i,1/\psi_i)\right)\,\,\CL(\spcurve_a) \right>_{\modsp_{g,n}} 
\eea

where
\beq\label{defCLspcurvea}
\CL(\spcurve_a) = \ee{\sum_k \hat t_{a,k} \kappa_k + {1\over 2} \sum_{\delta\in [\partial\modsp_{g,n}]}l_{\delta^*}\, \check B_{a,a}(1/\psi,1/\psi')}
\eeq
with the times $\hat t_{a,k}$ defined by the Laplace transform of $ydx$
\beq\label{defhattk}
\ee{-\sum_k \hat t_{a,k} u^{-k}} = \frac {u^{3/2}\,\ee{ux(a)}}{2\,\,\sqrt\pi}\, \int_{\gamma_a}\,ydx\,\ee{-ux}
\eeq
with $\gamma_a$ the steepest descent contour going through $a$, i.e. such that
\beq\label{defgammaa}
x(\gamma_a)-x(a)=\mathbb R_+,
\eeq
and $\hat B$ and $\check B$ are defined by Laplace transforms of the Bergman kernel
\beq\label{defhatBazu}
\hat B_{a}(z;u) = -\,\frac{\sqrt{u}\,\ee{ux(a)}}{\sqrt\pi}\,\int_{z'\in\gamma_a} B(z,z')\,\,\ee{-u x(z')} = \sum_d u^{-d}\,d\xi_{a,d}(z)
\eeq
\bea\label{defcheckBaauv}
\check B_{a,a}(u,v) &=&{uv\over u+v}+ \frac{\sqrt{uv}\,\ee{(u+v)x(a)}}{2\,\pi}\,\,\,\int_{z,z'\in\gamma_a} B(z,z')\,\,\ee{-u x(z)}\,\ee{-v x(z')} \cr
& =&  \sum_{k,l}\,\check B_{a,k;a,l}\,u^{-k}\,v^{-l} 
\eea
$[\partial\modsp_{g,n}]$ is the set of boundary divisors of $\modsp_{g,n}$, and $l_{\delta*}$ is the natural inclusion of a boundary of $\modsp_{g,n}$ into $\modsp_{g-1,n+2}\cup \cup^{\rm stable}_{h,m} \modsp_{h,m}\times\modsp_{g-h,n-m}$, and $\psi$, $\psi'$ represent the $\psi$ classes associated to the 2 marked points in 
$\modsp_{g-1,n+2}\cup \cup_{h,m} \modsp_{h,m}\times\modsp_{g-h,n-m}$, associated to the nodal point  in $\partial \modsp_{g,n}$.
\et

\proof{This theorem was proved in \cite{Eynard2011}, using the known result for the Kontsevich integral's spectral curve, and the deformation theory in \cite{EOFg} (special geometry) satisfied by symplectic invariants $W_n^{(g)}$'s. This allowed to view any spectral curve with one branch point, as a deformation of the Kontsevich's spectral curve, and thus prove this theorem.}

\medskip

Notice that formula (\ref{Wng1bpLaplace}) is very reminiscent of the ELSV formula \cite{EKEDAHL1999}.

\subsubsection{Example: topological vertex\label{sectopvertex}}

A very important example of application of theorem \ref{th1bp} is the topological vertex.

The spectral curve of the topological vertex is given by
\bea
\spcurve=(\mathbb C\setminus ]-\infty,0]\cup[1,\infty[,x,y,B) \cr
{} \cr
\left\{\begin{array}{l}
x(z) = -f\ln z -\ln{(1-z)} \cr
y(z)=-\ln z \cr
B(z_1,z_2) = \frac{dz_1\otimes dz_2}{(z_1-z_2)^2}
\end{array}\right.
\eea
It is more often described by observing that $X=\ee{-x}$ and $Y=\ee{-y}$ are related by the following algebraic equation (known as the mirror curve of $\mathbb C^3$ with framing $f$):
\beq
X=Y^f\,(1-Y).
\eeq

The only branchpoint is at $a=f/(1+f)$, and $\gamma_a=]0,1[$, so that the Laplace transform of $ydx$ is the Euler Beta function (we first integrate by parts):
\bea
\int_{\gamma_a} \,\ee{-xu}\,ydx 
&=& \frac 1 u \int_{\gamma_a} \ee{-xu}\,dy = \frac 1 u \int_0^1 z^{fu}\,(1-z)^u\,\frac{dz}{z} \cr
&=& \frac 1 u\,\, \frac{\Gamma(u+1)\Gamma(fu)}{\Gamma((f+1)u+1)} \cr
&=& \frac 1 {(f+1)u}\,\, \frac{\Gamma(u)\Gamma(fu)}{\Gamma((f+1)u)} \cr
\eea
Using the Stirling large $u$ expansion of the $\Gamma$ function, we thus have from \eqref{defhattk}
\beq
\ee{\hat t_0} = \sqrt{2f(f+1)},
\eeq
\beq
g(u) = \sum_{k\geq 1} \hat t_k u^{-k} = \sum_k \frac{\Ber_{2k}}{2k(2k-1)}\,u^{1-2k}\,\,\left(1+f^{1-2k}+(-f-1)^{1-2k}\right)
\eeq
where $\Ber_k$ is the $k^{\rm th}$ Bernoulli number. 
Similarly we find (see appendix \ref{appBuv})
\beq\label{Bcheckvertex}
\check B(u,v) = uv\,\,\frac{1-\ee{-g(u)}\,\,\ee{-g(v)}}{u+v} .
\eeq
Using a few combinatorial identities (see \cite{Eynard2011}), and thanks to Mumford's formula \eqref{eqmumfordhodge} (which writes Hodge classes in terms of $\psi$ and $\kappa$ classes, see \cite{Mumford1983}), we find that the spectral curve's class of the vertex is the product of 3 Hodge classes (see \cite{Eynard2011}):
\beq
\CL_a(\spcurve)\,\prod_{i=1}^n \ee{-g(1/\psi_i)} = \L_{\rm Hodge}(1)\L_{\rm Hodge}(f)\L_{\rm Hodge}(-f-1).
\eeq
Notice also that from \eqref{Bcheckvertex} we have
\beq
\int_{\gamma_a}\,\ee{-\mu x(z)}\,\, \hat B(z;1/\psi) =  {2\sqrt\pi\ee{-\mu x(a)}\over \sqrt\mu}\,\ee{-g(\mu)}\,\,\,{\mu\over 1+\mu\,\psi}\,\ee{-g(1/\psi)},
\eeq
i.e. formula \eqref{Wng1bp} reads in that case (after Laplace transform):
\bea
&& \int_{\gamma_a^n}\,\,W_n^{(g)}(\spcurve;z_1,\dots,z_n) \prod_i \ee{-\mu_i x(z_i)}  \cr
&=& 2^{d_{g,n}}\,\ee{\hat t_0(2g-2+n)}\,\,\int_{[\ovl\modsp_{g,n}]^{\rm vir}}\,\, \L_{\rm Hodge}(1)\L_{\rm Hodge}(f)\L_{\rm Hodge}(-f-1)\,\,  \prod_{i=1}^n {1\over 1+\mu_i\,\psi_i} \cr
&& \qquad \prod_{i=1}^n  {\Gamma(\mu_i)\Gamma(f\mu_i)\over \Gamma((f+1)\mu_i)}\left(\frac{(f+1)^{f+1}}{f^f}\right)^{\mu_i}\,\,\frac{2\sqrt{2\pi f}}{\sqrt{f+1}}\cr
\eea
whose right hand side is the famous Mari\~no--Vafa formula \cite{MV01}.

Using the result of \cite{ZhouJian2009, ChenLin2009}, i.e. that the symplectic invariants $W_n^{(g)}$ of the vertex are the Gromov-Witten invariants of $\mathbb C^3$, we see that \eqref{Wng1bp} translates into the Mari\~no--Vafa formula \cite{MV01}, and thus, $W_n^{(g)}$'s are generating functions of Gromov--Witten invariants of $\mathbb C^3$.

The large $f$ limit of that identity, is the ELSV formula \cite{EKEDAHL1999} combined with Bouchard--Mari\~no formula \cite{BM}.

\section{Spectral curve with several branchpoints}

Let $\spcurve=(\curve,x,y,B)$ be a spectral curve,
let $\{a_1,a_2,\dots,a_\bpt\}$ be the set of its branchpoints.
We first need to set up notations.

\subsection{Local description of the spectral curve near branchpoints}

For each branchpoint $a_i$ we define the steepest descent path $\gamma_{a_i}$, as a connected arc on $\curve$ passing through $a_i$ such that
\beq
x(\gamma_{a_i})-x(a_i)=\mathbb R_+ \, .
\eeq
In a vicinity of $a_i$ we define the local coordinate
\beq
\zeta_{a_i}(z) = \sqrt{x(z)-x(a_i)}.
\eeq

\br
For our purposes, it is sufficient that $\gamma_{a_i}$ is defined only in a vicinity of $a_i$, but for many examples, it is actually a well defined path in $\curve$.
\er

\subsubsection{Coefficients $\hat B_{a_i,k;a_j,l}$}

We expand the Bergman kernel in the vicinity of branchpoints as follows:
\beq\label{BTaylorexpaiaj}
B(z,z') \mathop{{\sim}}_{\stackrel{z'\to a_j}{z\to a_i}}\,\, \left(\frac{\delta_{i,j}}{(\zeta_{a_i}(z)-\zeta_{a_j}(z'))^2}
+ \sum_{d,d'\geq 0} B_{a_i,d;a_j,d'}\,\zeta_{a_i}(z)^d\,\zeta_{a_j}(z')^{d'}\,\right)\,\,d\zeta_{a_i}(z)\otimes d\zeta_{a_j}(z')
\eeq
and then we define
\beq
\hat B_{a_i,k;a_j,k'} = (2k-1)!!\,(2l-1)!!\,\,2^{-k-l-1}\,\,B_{a_i,2k;a_j,2k'}.
\eeq
It is useful to notice that the generating function of these last quantities can also be defined through Laplace transform, we define:
\beq
\check B_{a_i,a_j}(u,v)  =  \sum_{k,k'\geq 0}\, \hat B_{a_i,k;a_j,k'} u^{-k}\,v^{-l}, 
\eeq
which is given by the Laplace transform of the Bergman kernel
\beq
\check B_{a_i,a_j}(u,v) = \delta_{i,j}\,\frac{uv}{u+v} 
+ \frac{\sqrt{uv}\,\ee{ux(a_i)+vx(a_j)}}{2\pi}\,\int_{z\in\gamma_{a_i}}\,\int_{z'\in\gamma_{a_j}}\,B(z,z')\,\ee{-ux(z)}\,\ee{-vx(z')} 
\eeq
where the double integral is conveniently regularized when $i=j$, so that $\check B_{a_i,a_j}(u,v)$ is a power series of $u$ and $v$.

\subsubsection{Basis of differential forms $d\xi_{a_i,d}(z)$}

We define the set of functions $\xi_{a_i,d}(z)$ as follows:
\beq
d\xi_{a_i,d}(z) = -\,(2d-1)!!\,2^{-d}\,\Res_{z'\to a_i}\, B(z,z')\,\zeta_{a_i}(z')^{-2d-1}
\eeq
It is a meromorphic 1-form defined on $\curve$, with a pole only at $z=a_i$, of degree $2d+2$.

Namely, near $z\to a_j$  it behaves like
\beq
d\xi_{a_i,d}(z) \mathop{{\sim}}_{z\to a_j}\, -\,\delta_{i,j}\,\frac{(2d+1)!!\,d\zeta_{a_i}(z)}{2^d\,\zeta_{a_i}(z)^{2d+2}}\,-\frac{(2d-1)!!}{2^d}\,\sum_k B_{a_i,2d;a_j,k}\,\zeta_{a_j}(z)^k\,d\zeta_{a_j}(z).
\eeq

These differential forms will play an important role because they give the behavior of the Bergman kernel $B$ near a branchpoint:
\beq
B(z,z')-B(\bar z,z') \mathop{{\sim}}_{z\to a_i}\,  -2\,\sum_{d\geq 0}\,\,\frac{2^d}{(2d-1)!!}\,\,\zeta_{a_i}(z)^{2d}\,\,d\zeta_{a_i}(z)\otimes d\xi_{a_i,d}(z').
\eeq

$\xi_{a_j,0}(z)$ plays a special role, notice that it has a simple pole at $z=a_j$ and no other pole:
\beq
\xi_{a_j,0}(z) \mathop{{\sim}}_{z\to a_j} \frac{1}{\zeta_{a_j}(z)} + {\rm analytical}.
\eeq

\subsubsection{Laplace transform $f_{i,j}(u)$}

Knowing $\xi_{a_j,0}(z)$, it is useful to define its Laplace transform  along $\gamma_{a_i}$ as
\bea
f_{i,j}(u) 
&=& \, \frac{\sqrt u}{2\sqrt\pi}\,\ee{ux(a_i)}\,\, \int_{\gamma_{a_i}}\,\ee{-u\,x}\,\,\xi_{a_j,0}\,dx  \cr
&=& \, \frac{1}{2\sqrt{\pi\,u}}\,\ee{ux(a_i)}\,\, \int_{\gamma_{a_i}}\,\ee{-u\,x}\,\,d\xi_{a_j,0}  \cr
&=& \delta_{i,j}- \sum_{k\geq 0} \frac{\hat B_{a_j,0;a_i,k}}{u^{k+1}} .
\eea

In appendix \ref{appBuv}, we show that

\bl\label{lemmaBuv}[proved in appendix \ref{appBuv}]
If $\curve$ is a compact Riemann surface and $dx$ is a meromorphic form on $\curve$ and $B$ is the fundamental form of the second kind normalized on $\acycle$-cycles, we have
\beq
\check B_{a_i,a_j}(u,v) = \frac{uv}{u+v}\,\left( \delta_{i,j}- \sum_{k=1}^\bpt f_{i,k}(u)\,f_{j,k}(v)\right)
\eeq
\el
so that all we need to compute is in fact $f_{i,j}(u)$.

\subsubsection{Half Laplace transform}

We also define
\beq
B_{a_i}(z;u) = -\,\frac{\sqrt{u}\,\ee{u\,x(a_i)}}{\sqrt\pi}\,\,\int_{z'\in\gamma_{a_i}}\,B(z,z')\,\ee{-u\,x(z')} = \sum_d u^{-d}\,d\xi_{a_i,d}(z).
\eeq
If we do a second Laplace transform we have
\beq
\frac{\sqrt{v}\,\ee{v\,x(a_j)}}{2\sqrt\pi}\,\int_{z\in\gamma_{a_j}} B_{a_i}(z;u)\,\ee{-vx(z)} = \delta_{i,j}\,\frac{uv}{u+v}-\check B_{a_i,a_j}(v,u)  = \,\frac{uv}{u+v}\, \sum_k f_{i,k}(v)\,f_{j,k}(u).
\eeq

\subsubsection{The times $\hat t_{a_i,k}$}

Finally we define the times $\hat t_{a_i,k}$ at branchpoint $a_i$ in terms of the local behavior of $y(z)$ by the Laplace transform of $ydx$ along $\gamma_{a_i}$
\beq
\ee{-\hat t_{a_i,0}}\,\ee{-g_{a_i}(u)} 
= \frac{u^{3/2}\,\ee{ux(a_i)}}{2\sqrt{\pi}}\,\int_{z\in\gamma_{a_i}}\,\, \ee{-ux(z)}\,y(z)\,dx(z)
= \frac{\sqrt{u}\,\ee{ux(a_i)}}{2\sqrt{\pi}}\,\int_{z\in\gamma_{a_i}}\,\, \ee{-ux(z)}\,dy(z)
\eeq
The times $\hat t_{a_i,k}$ are the coefficients of the expansion of $g(u)$ at large $u$:
\beq
g_{a_i}(u) = \sum_{k\geq 1} \hat t_{a_i,k} u^{-k}.
\eeq
Notice that the time $\hat t_{a_i,0}$ is given by
\beq
\ee{-\hat t_{a_i,0}}  = \frac{1}{4}\,\lim_{z\to a_i}\,\frac{y(z)-y(\bar z)}{\zeta_{a_i}(z)} = \frac{y'(a_i)}{\sqrt{2\,x''(a_i)}}.
\eeq

\subsection{Structure of invariants \label{secstructWng}}

Since the definition of $W_n^{(g)}$ involves only residues at branchpoints, this means that for each $(g,n)$, $W_n^{(g)}$ is a polynomial of the coefficients of $B$ and $y$ in a Taylor expansion near the branchpoints, in other words, $W_n^{(g)}$ is a polynomial in the times $\hat t_{a_i,k}$ and $B_{a_i,k;a_j,l}$.

Since $K$ is linear in $B$, and $W_2^{(0)}=B$ is also linear in $B$, one easily finds by recursion that $W_n^{(g)}$ is polynomial in the coefficients $B_{a_i,d;a_j,d'}$, of degree $3g-3+2n=d_{g,n}+n$.

When we compute residues at $z\to a_i$, we may replace $B(z,z')$, or more precisely the combination  $B(z,z')-B(\bar z,z')$ (indeed the Bergmann kernels always enter the residue computation with this combination, this due to the fact that $K(z_0,z)=K(z_0,\bar z)$) by 
\beq
B(z,z')-B(\bar z,z') = -2\,\sum_{d} \zeta_{a_i}(z)^{2d}\,d\zeta_{a_i}(z)\,\,\,\frac{2^d}{(2d-1)!!}\, d\xi_{a_i,d}(z'),
\eeq
where
\beq
d\xi_{a_i,d}(z) = -\frac{(2d-1)!!}{2^d}\,\Res_{z'\to a_i}\,B(z,z')\,\, \zeta_{a_i}(z)^{-2d-1}
\eeq
and thus one clearly sees by recursion that there exists some coefficients $A_n^{(g)}$ such that
\beq
W_n^{(g)}(\spcurve; z_1,\dots,z_n)  = \sum_{i_1,\dots,i_n}\sum_{d_1,\dots,d_n}\,A_n^{(g)}(\spcurve; i_1,d_1;\dots;i_n,d_n)\,\prod_{k=1}^n d\xi_{a_{i_k},d_k}(z_k)
\eeq
and the coefficients $A_n^{(g)}$ are polynomials of the $\hat B_{a_i,d;a_j,d'}$ of degree $3g-3+n=d_{g,n}$. $A_n^{(g)}$ are also polynomials in the $\hat t_{a_i,k}$ because the residues also picks terms in the Taylor expansion of the denominator of $K$, i.e. the Taylor expansion of $y(z)dx(z)$ near branchpoints, i.e. the coefficients $\hat t_{a_i,k}$.

So we have:

\bp
For any spectral curve $\spcurve$ with branchpoints $a_1,\dots,a_\bpt$, there exists some coefficients
$A_n^{(g)}(\spcurve; i_1,d_1;\dots;i_n,d_n)$ such that
\beq
W_n^{(g)}(\spcurve; z_1,\dots,z_n)  =  \sum_{i_1,\dots,i_n}\sum_{d_1,\dots,d_n}\,A_n^{(g)}(\spcurve; i_1,d_1;\dots;i_n,d_n)\,\prod_{k=1}^n d\xi_{a_{i_k},d_k}(z_k)
\eeq

- The coefficients $A_n^{(g)}$ are non--vanishing only if $\sum_k d_k\leq 3g-3+n$.

- The coefficients $A_n^{(g)}$ are polynomials of the $\hat B_{a_i,d;a_j,d'}$ of degree $3g-3+n=d_{g,n}$. 

- The coefficients $A_n^{(g)}$ are polynomials in the $\hat t_{a_i,k}$, of degree at most $d_{g,n}$ (where $\hat t_{a_i,k}$ is weighted with degree $k$).

\ep

Our goal now, is to show that the coefficients $A_n^{(g)}$ can be written as intersection numbers in some moduli space.

\subsection{Definition of an appropriate moduli space}

\bd
We define
\beq
\ovl\modsp_{g,n}^\bpt  = \left\{ (\Sigma;p_1,\dots,p_n;\sigma) \right\}
\eeq
where $ (\Sigma;p_1,\dots,p_n) \in \ovl\modsp_{g,n}$ is a Deligne--Mumford stable curve of genus $g$ with $n$ marked points, and $\sigma:\Sigma\setminus \{{\rm nodal\,points}\}\to \{1,2,\dots,\bpt\}$ is a continuous map.

\ed

This moduli space is clearly compact, and contains $\bpt$ copies of $\ovl\modsp_{g,n}$.

Notice that if $ (\Sigma;p_1,\dots,p_n)$ is a smooth curve ($\in \modsp_{g,n}$), i.e. with no nodal points then $\sigma$ must be constant, and if $(\Sigma;p_1,\dots,p_n)$ is a nodal curve with $k$ components, then $\sigma$ must be piecewise constant i.e. is constant on each component.

\bd\label{defLMnbeta}
Let $\{a_1,a_2,\dots,a_\bpt\}$ be a set of $\bpt$ elements (later on, these will be the branchpoints of some spectral curve).
For every $i=1,\dots, \bpt$, let $\hat\Lambda_{a_i}=f_i(\psi_1,\dots,\psi_n,\kappa_0,\kappa_1,\kappa_2,\dots)$ be a tautological class on $\ovl\modsp_{g,n}$, which is a combination of $\psi$ and $\kappa$ classes, defined independently of $g$ and $n$ (as for instance \eqref{defCLspcurvea}).
Let $C_{a_i,d;a_j,d'}$ be an arbitrary sequence of complex numbers indexed by $i,j\in [1,\dots,\bpt]$ and $d,d'\in \mathbb N$.
We define the class $\hat\L$ on $\ovl\modsp_{g,n}^\bpt$ as follows:

If $(\Sigma;p_1,\dots,p_n;\sigma)\in \ovl\modsp_{g,n}^\bpt$ is such that $(\Sigma;p_1,\dots,p_n)$ is a stable map with $k$ components $\cup_{m=1}^k (\Sigma_m, \vec p_m, \vec q_m)$ where $\vec p_m$ is a subset of $\{p_1,\dots,p_n\}$ and pairs $(q_i,q_j)$ are the nodal points (see fig.\ref{fignodalmap1}), we define the topological class
\beq
\hat\L = \prod_{m=1}^k \hat\L_{a_{\sigma(\Sigma_m)}} \prod_{<q_i,q_j>={\rm nodal\,point}} \left(\sum_{d,d'} C_{a_{\sigma(q_i)},d;a_{\sigma(q_j)},d'}\,l_*\,\psi(q_i)^d\psi(q_j)^{d'}\right)
\eeq
where $l_*$ is the natural inclusion into $\modsp_{g_{m},n_{m}}\times \modsp_{g_{m'},n_{m'}}$ (if $q_i\in \Sigma_m$ and $q_j\in\sigma_{m'}$).
\ed

Then we have our main theorem:
\bt\label{mainth}
For any spectral curve $\spcurve=(\curve,x,y,B)$ with $\bpt$ branchpoints $\{a_1,a_2,\dots,a_\bpt\}$, 
we consider $\spcurve_{a_i} = (\curve_i,x,y,\Brond)$ where $\curve_i\subset\curve$ is a vicinity of the branchpoint $a_i$, $x$ and $y$ are the restrictions of $x$ and $y$ to $\curve_i$, and $\Brond$ can be any arbitrary Bergman kernel chosen in $\curve_i$ (it doesn't need to be the restriction of $B$ to $\curve_i\times \curve_i$).
\smallskip

Then, the symplectic invariants of $\spcurve$ are integrals of  the class $\hat\L(\spcurve)$ on $\ovl\modsp_{g,n}^\bpt$:
\beq
\encadremath{
W_n^{(g)}(\spcurve;z_1,\dots,z_n) = 2^{d_{g,n}}\,\int_{\ovl\modsp_{g,n}^\bpt} \hat\L(\spcurve)\,\, \prod_{i=1}^n \hat B_{a_{\sigma(p_i)}}(z_i;1/\psi(p_i))
}\eeq
or in Laplace transform
\beq
\encadremath{
\begin{array}{l}
\displaystyle \quad \int_{z_i\in \gamma_{a_{k_i}}}\,\,W_n^{(g)}(\spcurve;z_1,\dots,z_n)\,\,\prod_{i=1}^n\ee{-\mu_ix(z_i)} \cr
\displaystyle = 2^{d_{g,n}}\,\prod_{i=1}^n \frac{2\sqrt{\pi}\,\ee{-\mu_i\,x(a_{k_i})}}{\sqrt{\mu_i}}\,\int_{\ovl\modsp_{g,n}^\bpt} \hat\L(\spcurve)\,\, \prod_{i=1}^n \left(  \frac{\mu_i\,\delta_{k_i,\sigma(p_i)}}{1+\mu_i\psi(p_i)} - \check B_{a_{k_i},a_{\sigma(p_i)}}(\mu_i,1/\psi(p_i))\right)
\end{array}
}\eeq
or using lemma \ref{lemmaBuv}
\beq
\encadremath{
\begin{array}{l}
\displaystyle \quad \int_{z_i\in \gamma_{a_{k_i}}}\,\,W_n^{(g)}(\spcurve;z_1,\dots,z_n)\,\,\prod_{i=1}^n\ee{-\mu_ix(z_i)} \cr
\displaystyle = 2^{d_{g,n}}\,\prod_{i=1}^n (2\sqrt{\pi\mu_i}\,\ee{-\mu_i\,x(a_{k_i})})\,\int_{\ovl\modsp_{g,n}^\bpt} \hat\L(\spcurve)\,\, \prod_{i=1}^n \frac{\sum_{r=1}^\bpt  f_{k_i,r}(\mu_i) f_{\sigma(p_i),r}(1/\psi(p_i))}{1+\mu_i\psi(p_i)}
\end{array}
}\eeq

In particular for $n=0$:
\beq
\encadremath{
F_g(\spcurve) =  2^{d_{g,0}}\,\int_{\ovl\modsp_{g,0}^\bpt} \hat\L(\spcurve)
}\eeq
where $\hat\L(\spcurve)$ is defined as in def.\ref{defLMnbeta}, with $\hat\L_{a_i}$ defined in theorem \ref{th1bp}:
 if $(\Sigma,p_1,\dots,p_n,\sigma)\in\modsp_{g,n}^\bpt$ is made of $k$ components $\Sigma=\cup_{m=1}^k \Sigma_m$, with $\vec p_m = \{p_1,\dots,p_n\}\cap\Sigma_m$ and $(q_i,q_j)$ are the nodal points, we define
\bea
\hat\Lambda(\spcurve) 
&=& \prod_{m=1}^k \hat\Lambda_{a_m}\,\,
\prod_{<q_i,q_j>={\rm nodal\,point}}\,\big(\sum_{d,d'} (\hat B_{a_{\sigma(q_i)},d;a_{\sigma(q_j)},d'} \cr 
&&\qquad\qquad -\delta_{\sigma(q_i),\sigma(q_j)}\hat \Brond_{a_{\sigma(q_i)},d;a_{\sigma(q_i)},d'})\,l_* \psi(q_i)^d\,\psi(q_j)^{d'} \big) \cr
&=& \prod_{m=1}^k \hat\Lambda_{a_m}\,\,\prod_{<q_i,q_j>={\rm nodal\,point}}\,\big(\check B_{a_{\sigma(q_i)},a_{\sigma(q_j)}}(1/\psi(q_i),1/\psi(q_j))  \cr
&& \qquad \quad - \delta_{\sigma(q_i),\sigma(q_j)}\,\check \Brond_{a_{\sigma(q_i)},a_{\sigma(q_i)}}(1/\psi(q_i),1/\psi(q_j))\big) , \cr
\eea
where $\hat\L_{a_i}$ is the class defined in theorem \ref{th1bp} or in \cite{Eynard2011}:
\beq
\CL_{a_i} = \ee{\sum_{k} \hat t_{a_i,k}\,\kappa^k +\frac{1}{2}\sum_{\delta\in\partial\modsp_{g,n}}l_{\delta^*}\,\check \Brond_{a_i,a_i}(1/\psi,1/\psi')}
\eeq
with the times $\hat t_{a_i,k}$ defined by the Laplace transform of $ydx$ along $\gamma_{a_i}$
\beq
\ee{-\sum_k \hat t_{a_i,k} u^{-k}} = \frac{u\sqrt{u}\,\,\ee{ux(a_i)}}{2\sqrt\pi}\,\,\int_{z\in\gamma_{a_i}} \ee{-ux(z)}\,\,y(z)dx(z),
\eeq
and where the $\hat B_{a_i,d;a_j,d'}$ are defined by the Laplace transform of the Bergman kernel
\bea
\check B_{a_i,a_j}(u,v) 
&=& \delta_{i,j}\,\frac{uv}{u+v}+\frac{\sqrt{uv}\,\,\ee{u\,x(a_i)+vx(a_j)}}{2\pi}\,\,\int_{z\in\gamma_{a_i}}\int_{z'\in\gamma_{a_j}}\,\,B(z,z')\,\,\ee{-u\,x(z)}\,\,\ee{-v\,x(z')}  \cr
&=& \sum_{d,d'}\,\hat B_{a_i,d;a_j,d'}\,\,u^{-d}\,\,v^{-d'} \cr
&=& \frac{uv}{u+v}\,\left(\delta_{i,j}-\sum_{r=1}^\bpt f_{i,r}(u)\,f_{j,r}(v)\right)
\eea
and
\bea
\hat B_{a_i}(z;u) 
&=&  -\,\frac{\sqrt{u}\,\,\ee{u\,x(a_i)}}{\sqrt\pi}\,\,\int_{z'\in\gamma_{a_j}}\,\,B(z,z')\,\,\ee{-u\,x(z)}  \cr
&=& \sum_d u^{-d}\,\,d\xi_{a_i,d}(z),
\eea
and
\beq
f_{i,j}(u) = \frac{\sqrt{u}\,\ee{ux(a_i)}}{2\sqrt{\pi}}
\,\,\int_{z\in \gamma_{a_i}}\,
\ee{-u x(z)}\,\xi_{a_j,0}(z)\,dx(z)
\eeq
\et

\proof{The proof almost follows the definition of $W_n^{(g)}$'s. It can also be viewed as a simple application of the method of Kostov and Orantin \cite{Kostov1999, Kostov2010,OrantinN.2008}. We give the full proof in appendix \ref{appproof}.}



\subsection{How to use the formula}

Example, assume that there are 2 branch points $a_1, a_2$, and  let us compute $W_3^{(0)}$ and $W_4^{(0)}$.

\smallskip 
$\bullet$ For $W_3^{(0)}$, consider a stable curve $\Sigma$ of genus 0, with 3 marked points $(p_1,p_2,p_3)$. The only possibility is that $\Sigma$ is a sphere with 3 marked points, and thus it is a smooth surface, and $\sigma$ must be constant. There are two possibilities $\sigma=1$ or $\sigma=2$. In other words $\modsp_{0,3}^2 \sim \modsp_{0,3}\cup \modsp_{0,3}$.
We thus have
\bea
W_3^{(0)}(\spcurve,z_1,z_2,z_3) 
&=& \left< \CL_1\,\,\hat B_{a_1}(z_1,1/\psi_1)\,\hat B_{a_1}(z_2,1/\psi_2)\,\hat B_{a_1}(z_3,1/\psi_3)\,\right>_{0,3} \cr
&& + \left< \CL_2\,\,\hat B_{a_2}(z_1,1/\psi_1)\,\hat B_{a_2}(z_2,1/\psi_2)\,\hat B_{a_2}(z_3,1/\psi_3)\,\right>_{0,3} .
\eea
Moreover, we have $\hat B_{a_i}(z,1/\psi) = \sum_d \psi^d\,\, d\xi_{a_i,d}(z)$, and since $\dim \modsp_{0,3}=d_{0,3}=0$, only the term $d=0$ may contribute, i.e.
\bea
W_3^{(0)}(\spcurve,z_1,z_2,z_3) 
&=& \left< \CL_1\,\right>_{0,3}\,d\xi_{a_1,0}(z_1)\,d\xi_{a_1,0}(z_2)\,d\xi_{a_1,0}(z_3) \cr
&& + \left< \CL_2\,\right>_{0,3}\,d\xi_{a_2,0}(z_1)\,d\xi_{a_2,0}(z_2)\,d\xi_{a_2,0}(z_3). 
\eea
Then, we have
\beq
\CL_1 = \ee{\sum_k \hat t_{a_1,k}\kappa_k}\,\,\ee{\frac{1}{2}\,\sum_{\delta\in \d\modsp_{0,3}} \sum_{d,d'} l_{\delta*}\,\hat {\Brond_1}_{a_1,d;a_1,d'}\psi^d\,\psi'^{d'}}
\eeq
and since $\d\modsp_{0,3}=\emptyset$, and since $d_{0,3}=0$, only the term $\kappa_0$ may contribute, and thus
\beq
\left<\CL_1\right>_{0,3} = \ee{\hat t_{a_1,0}}\,\left<1\right>_{0,3} = \ee{\hat t_{a_1,0}}
\virg
\left<\CL_2\right>_{0,3} = \ee{\hat t_{a_2,0}}\,\left<1\right>_{0,3} = \ee{\hat t_{a_2,0}},
\eeq
and finally, since $<1>_{0,3}=1$:
\bea
W_3^{(0)}(\spcurve,z_1,z_2,z_3) 
&=& \ee{\hat t_{a_1,0}}\,d\xi_{a_1,0}(z_1)\,d\xi_{a_1,0}(z_2)\,d\xi_{a_1,0}(z_3) \cr
&& + \ee{\hat t_{a_2,0}}\,d\xi_{a_2,0}(z_1)\,d\xi_{a_2,0}(z_2)\,d\xi_{a_2,0}(z_3) .
\eea

\medskip
$\bullet$ For $W_4^{(0)}$, consider a stable curve $\Sigma$ of genus 0, with 4 marked points $(p_1,p_2,p_3,p_4)$.
If $\Sigma$ is smooth, then a continuous map $\sigma:\Sigma\to \{1,2\}$ must be constant, it takes either the value $\sigma=1$ or $\sigma=2$.
If $\Sigma$ is not smooth, then it is a nodal curve, let us call $(q_1,q_2)$ the nodal point. The only stable possibility is that $\Sigma$ has 2 components 
$\Sigma=\Sigma_1\cup\Sigma_2$, which are both spheres with 3 marked points,
and thus $q_1\in\Sigma_1$, $q_2\in\Sigma_2$, and 2 of the points $p_1,p_2,p_3,p_4$ belong to $\Sigma_1$ and the 2 others belong to $\Sigma_2$.
Then a continuous map $\sigma:\Sigma\to \{1,2\}$ must be constant on $\Sigma_1$ and on $\Sigma_2$, for instance it may take the value $\sigma=1$ on $\Sigma_1$ and  $\sigma=2$ on $\Sigma_2$.
This shows that
\beq
\modsp_{0,4}^2 \sim \modsp_{0,4}\,\cup\, \modsp_{0,4}\, \cup\, \overbrace{\modsp_{0,3}\times \modsp_{0,3}}^{12\,{\rm times}} ,
\eeq
where the $12$ times correspond to the 12 possibilities to have 2 point with color 1 and 2 point with color 2 among the 4 marked points $p_1,p_2,p_3,p_4$.

We thus have
\bea
&& \frac{1}{2}\,W_4^{(0)}(\spcurve, z_1,\dots,z_4) \cr
&=&  \left<\prod_{i=1}^4 \hat B_{a_1}(z_i,1/\psi(p_i))\,\, \CL_1 \right>_{0,4} \cr
&& + \left<\prod_{i=1}^4 \hat B_{a_2}(z_i,1/\psi(p_i))\,\, \CL_2 \right>_{0,4} \cr
&& + \frac{1}{2}\sum_{i,j\in \{1,2\}^2}\sum_{d,d'} C_{a_i,d;a_j,d'}\,\, \left<\hat B_{a_i}(z_1,1/\psi(p_1))\,\hat B_{a_i}(z_2,1/\psi(p_2))\,\, \psi(q_1)^d\,\, \CL_i \right>_{0,3}\,\cr
&& \qquad \qquad \qquad \qquad \left<\hat B_{a_j}(z_3,1/\psi(p_3))\,\hat B_{a_j}(z_4,1/\psi(p_4))\,\, \psi(q_2)^{d'}\,\, \CL_j \right>_{0,3} \cr
&& + {\rm symmetrize\,on\,}(z_1,z_2,z_3,z_4)
\eea 
where all intersection numbers in the right hand side are now usual intersection numbers in the corresponding  $\ovl\modsp_{g,n}$.

Since $\modsp_{0,3}$ is a point ($\dim \modsp_{0,3}=d_{0,3}=0$), we must have $d=d'=0$ in the last term, and we may replace $\hat B_{a_i}(z,1/\psi) = \sum_d \psi^d\,d\xi_{a_i,d}(z)$ by $d\xi_{a_i,0}(z)$.
And since $d_{0,4}=1$, we may replace $\hat B_{a_i}(z,1/\psi) = \sum_d \psi^d\,d\xi_{a_i,d}(z)$ by $d\xi_{a_i,0}(z)+\psi\,d\xi_{a_i,1}(z)$ in the $<>_{0,4}$ terms.
More explicitly, that gives
\bea
&& \frac{1}{2}\,W_4^{(0)}(\spcurve, z_1,\dots,z_4) \cr
&=&  \left<\prod_{i=1}^4 (d\xi_{a_1,0}(z_i)+\psi(p_i)\,d\xi_{a_1,1}(z_i))\,\, \CL_1 \right>_{0,4} \cr
&& +  \left<\prod_{i=1}^4 (d\xi_{a_2,0}(z_i)+\psi(p_i)\,d\xi_{a_2,1}(z_i))\,\, \CL_2 \right>_{0,4} \cr
&& + \frac{1}{2}\sum_{i,j\in \{1,2\}^2} C_{a_i,0;a_j,0}\,\, \left< \CL_i \right>_{0,3}\,d\xi_{a_i,0}(z_1)\,d\xi_{a_i,0}(z_2)\,\, \left< \CL_j \right>_{0,3}\,d\xi_{a_j,0}(z_3)\,d\xi_{a_j,0}(z_4)\cr
&& + {\rm symmetrize\,on\,}(z_1,z_2,z_3,z_4)
\eea 
Again, since $d_{0,3}=0$, we may keep only the $\kappa_0$ term in the computation of $<\CL_i>_{0,3}$, i.e.
\beq
\left< \CL_i \right>_{0,3} = \left< \ee{\hat t_{a_i,0} \kappa_0} \right>_{0,3}  =\ee{\hat t_{a_i,0}}.
\eeq
Since $d_{0,4}=1$, we need to keep $\kappa_0$ and $\kappa_1$, and since $\d \modsp_{0,4}\sim \overbrace{\modsp_{0,3}\times \modsp_{0,3}}^{6\,{\rm times}}$, we have
\bea
\left< \CL_i \right>_{0,4} 
&=& \left< \ee{\hat t_{a_i,0} \kappa_0}\,(1+\hat t_{a_i,1} \kappa_1) \right>_{0,4}  + \frac{6}{2} \hat \Brond_{a_i,0;a_i,0}\,\left< \CL_i \right>_{0,3}\,\left< \CL_i \right>_{0,3} \cr
&=& \hat t_{a_i,1}\,\left< \ee{\hat t_{a_i,0} \kappa_0}\, \kappa_1 \right>_{0,4}  + 3 \hat \Brond_{a_i,0;a_i,0}\,\,\,\ee{\hat t_{a_i,0}}\,\,\ee{\hat t_{a_i,0}} \cr
&=& \hat t_{a_i,1}\,\ee{2\hat t_{a_i,0}}\,\,\,\left< \kappa_1 \right>_{0,4}  + 3 \hat \Brond_{a_i,0;a_i,0}\,\,\,\ee{\hat t_{a_i,0}}\,\,\ee{\hat t_{a_i,0}} \cr
&=& \ee{2\hat t_{a_i,0}}\,\,\,\left( \hat t_{a_i,1}\,   + 3 \hat \Brond_{a_i,0;a_i,0}\right) 
\eea
where we used that $<\kappa_1>_{0,4}=1$.
Similarly, since $\dim \psi=1$, we have
\beq
\left< \psi \CL_i \right>_{0,4} 
= \left< \psi\,\ee{\hat t_{a_i,0} \kappa_0} \right>_{0,4}
= \,\ee{2\hat t_{a_i,0}}\,\,\left< \psi \right>_{0,4}= \,\ee{2\hat t_{a_i,0}}\,\,\left< \tau_1 \right>_{0,4}
= \,\ee{2\hat t_{a_i,0}},
\eeq
where we used that $<\tau_1>_{0,4}=1$.
We thus have
\bea
&& \frac{1}{2}\,W_4^{(0)}(\spcurve, z_1,\dots,z_4) \cr
&=&  \ee{2\hat t_{a_1,0}}\,\,\left( \hat t_{a_1,1}\,   + 3\, \hat \Brond_{a_1,0;a_1,0}\right)\,d\xi_{a_1,0}(z_1)\,d\xi_{a_1,0}(z_2)\,d\xi_{a_1,0}(z_3)\,d\xi_{a_1,0}(z_4) \cr
&& +  \ee{2\hat t_{a_1,0}}\,\Big(\,d\xi_{a_1,1}(z_1)\,d\xi_{a_1,0}(z_2)\,d\xi_{a_1,0}(z_3)\,d\xi_{a_1,0}(z_4)  + {\rm sym.\,on\,}(z_1,z_2,z_3,z_4) \Big)\cr
&& +  \ee{2\hat t_{a_2,0}}\,\,\left( \hat t_{a_2,1}\,   + 3\, \hat \Brond_{a_2,0;a_2,0}\right)\,d\xi_{a_2,0}(z_1)\,d\xi_{a_2,0}(z_2)\,d\xi_{a_2,0}(z_3)\,d\xi_{a_2,0}(z_4) \cr
&& +  \ee{2\hat t_{a_2,0}}\,\Big(\,d\xi_{a_2,1}(z_1)\,d\xi_{a_2,0}(z_2)\,d\xi_{a_2,0}(z_3)\,d\xi_{a_2,0}(z_4)  + {\rm sym.\,on\,}(z_1,z_2,z_3,z_4) \Big)\cr
&& \cr
&& + \frac{1}{2}\sum_{i,j\in \{1,2\}^2} C_{a_i,0;a_j,0}\,\, \ee{\hat t_{a_i,0}+\hat t_{a_j,0}}\,\,\Big(\,\,d\xi_{a_i,0}(z_1)\,d\xi_{a_i,0}(z_2) \,\,\,d\xi_{a_j,0}(z_3)\,d\xi_{a_j,0}(z_4)\cr
&& + {\rm sym.\,on\,}(z_1,z_2,z_3,z_4) \Big)\cr
&=&  \ee{2\hat t_{a_1,0}}\,\,\left( \hat t_{a_1,1}\,   + 3\, \hat \Brond_{a_1,0;a_1,0}\right)\,d\xi_{a_1,0}(z_1)\,d\xi_{a_1,0}(z_2)\,d\xi_{a_1,0}(z_3)\,d\xi_{a_1,0}(z_4) \cr
&& +  \ee{2\hat t_{a_1,0}}\,\Big(\,d\xi_{a_1,1}(z_1)\,d\xi_{a_1,0}(z_2)\,d\xi_{a_1,0}(z_3)\,d\xi_{a_1,0}(z_4)  + {\rm sym.\,on\,}(z_1,z_2,z_3,z_4) \Big)\cr
&& +  \ee{2\hat t_{a_2,0}}\,\,\left( \hat t_{a_2,1}\,   + 3\, \hat \Brond_{a_2,0;a_2,0}\right)\,d\xi_{a_2,0}(z_1)\,d\xi_{a_2,0}(z_2)\,d\xi_{a_2,0}(z_3)\,d\xi_{a_2,0}(z_4) \cr
&& +  \ee{2\hat t_{a_2,0}}\,\Big(\,d\xi_{a_2,1}(z_1)\,d\xi_{a_2,0}(z_2)\,d\xi_{a_2,0}(z_3)\,d\xi_{a_2,0}(z_4)  + {\rm sym.\,on\,}(z_1,z_2,z_3,z_4) \Big)\cr
&& \cr
&& + \frac{1}{2}\, C_{a_1,0;a_1,0}\,\, \ee{2\hat t_{a_1,0}}\,\,\Big( 6\,\,d\xi_{a_1,0}(z_1)\,d\xi_{a_1,0}(z_2) \,d\xi_{a_1,0}(z_3)\,d\xi_{a_1,0}(z_4) \Big)\cr
&& + \frac{1}{2}\, C_{a_2,0;a_2,0}\,\, \ee{2\hat t_{a_2,0}}\,\,\Big( 6\,\,d\xi_{a_2,0}(z_1)\,d\xi_{a_2,0}(z_2) \,d\xi_{a_2,0}(z_3)\,d\xi_{a_2,0}(z_4) \Big)\cr
&& +  C_{a_1,0;a_2,0}\,\, \ee{\hat t_{a_1,0}+\hat t_{a_2,0}}\,\,\Big(\,\,d\xi_{a_1,0}(z_1)\,d\xi_{a_1,0}(z_2) \,\,\,d\xi_{a_2,0}(z_3)\,d\xi_{a_2,0}(z_4)\cr
&& + {\rm sym.\,on\,}(z_1,z_2,z_3,z_4) \Big)\cr
\eea 
and then, remember that
\beq
C_{a_i,0;a_j,0} = \hat B_{a_i,0;a_j,0}-\delta_{i,j}\,\hat \Brond_{a_i,0;a_j,0}
\eeq
i.e. finally:
\bea
&& \frac{1}{2}\,W_4^{(0)}(\spcurve, z_1,\dots,z_4) \cr
&=&  \ee{2\hat t_{a_1,0}}\,\,\left( \hat t_{a_1,1}\,   + 3\, \hat B_{a_1,0;a_1,0}\right)\,d\xi_{a_1,0}(z_1)\,d\xi_{a_1,0}(z_2)\,d\xi_{a_1,0}(z_3)\,d\xi_{a_1,0}(z_4) \cr
&& +  \ee{2\hat t_{a_1,0}}\,\Big(\,d\xi_{a_1,1}(z_1)\,d\xi_{a_1,0}(z_2)\,d\xi_{a_1,0}(z_3)\,d\xi_{a_1,0}(z_4)  + {\rm symmetrize\,on\,}(z_1,z_2,z_3,z_4) \Big)\cr
&& +  \ee{2\hat t_{a_2,0}}\,\,\left( \hat t_{a_2,1}\,   + 3\, \hat B_{a_2,0;a_2,0}\right)\,d\xi_{a_2,0}(z_1)\,d\xi_{a_2,0}(z_2)\,d\xi_{a_2,0}(z_3)\,d\xi_{a_2,0}(z_4) \cr
&& +  \ee{2\hat t_{a_2,0}}\,\Big(\,d\xi_{a_2,1}(z_1)\,d\xi_{a_2,0}(z_2)\,d\xi_{a_2,0}(z_3)\,d\xi_{a_2,0}(z_4)  + {\rm symmetrize\,on\,}(z_1,z_2,z_3,z_4) \Big)\cr
&& \cr
&& +  \hat B_{a_1,0;a_2,0}\,\, \ee{\hat t_{a_1,0}+\hat t_{a_2,0}}\,\,\Big(\,\,d\xi_{a_1,0}(z_1)\,d\xi_{a_1,0}(z_2) \,\,\,d\xi_{a_2,0}(z_3)\,d\xi_{a_2,0}(z_4)\cr
&& + {\rm symmetrize\,on\,}(z_1,z_2,z_3,z_4) \Big).
\eea 
It is easy to verify that this coincides with the direct computation of residues in the very definition of $W_4^{(0)}$.

\subsection{Examples}

In this section, we consider some classical examples of spectral curves with 2 branchpoints, and we compute  the corresponding Laplace transforms.

\subsubsection{Ising model and the Airy function}

The spectral curve of the Ising model coupled to gravity, i.e. the $(4,3)$ minimal model, is (see \cite{Eynard2008}):
\bea
\spcurve=({\mathbb C},x,y,B) \cr
{} \cr
\left\{\begin{array}{l}
x(z)=z^3-3z \cr
y(z)=z^4-4z^2+2 \cr
B(z_1,z_2) = \frac{dz_1\otimes dz_2}{(z_1-z_2)^2}
\end{array}\right.
\eea
That curve has two branchpoints:
\beq
a_+=1 \virg a_-=-1.
\eeq
Instead of chosing the colors $\sigma\in \{1,2\}$, we chose to name the colors $\sigma\in \{+,-\}$.

Notice that we have the symmetry
\beq
x(-z)=-x(z) \virg y(-z)=y(z),
\eeq
and thus all what we compute for $a_+$ can easily be transposed to $a_-$.

\bigskip

We have:
\beq
\zeta_+(z)=\sqrt{x(z)+2} = (z-1)\sqrt{z+2}.
\eeq
We easily find
\beq
\xi_{a_+,0}(z) = \frac{1}{\sqrt 3}\,\frac{1}{z-1} 
= \frac{1}{\zeta_+} + \frac{3}{2}\,\sum_k\,\,\frac{(-1)^k\,\zeta_+^k}{3^{3/2(k+1)}}\,\,\frac{\Gamma(3k/2)}{(k+1)!\,\Gamma(k/2)}
\eeq
i.e.
\beq
B_{+,0;+,k} = \,\,\frac{(-1)^{k}}{2\,\,3^{3k/2+2}}\,\,\frac{\Gamma(3/2+3k/2)}{(k+2)!\,\Gamma(1/2+k/2)}
\eeq
\beq
\hat B_{+,0;+,k} 
= \frac{1}{2^{3(k+1)}\,\,3^{3k+2}}\,\,\frac{(6k+1)!!}{(2k+2)!\,}
= \frac{1}{2^{4(k+1)}\,\,3^{3k+2}}\,\,\frac{(6k+1)!!}{(k+1)!\,\,(2k+1)!!\,}
\eeq
%
i.e.
\beq
f_{+,+}(u) = 1- \sum_k \frac{1}{2^{4k}\,3^{3k-1}}\,\,\frac{(6k-5)!!}{k!\,(2k-1)!!\,}\,u^{-k}  
\eeq
Since $f_{+,+}$ can also be obtained from the Laplace transform, we write
\bea
f_{+,+}(u^{3/2}/3) 
&=& \frac{(u^{3/2}/3)^{1/2}\,\ee{-\frac{2}{3}\,u^{3/2}}}{2\sqrt{\pi}} \int \ee{-u^{3/2}/3(z^3-3z)}\,\frac{1}{\sqrt{3}\,(z-1)}\,\,3(z^2-1)\,dz  \cr
&=& \frac{u^{3/4}\,\ee{-\frac{2}{3}\,u^{3/2}}}{2\sqrt{\pi}} \int \ee{-u^{3/2}/3(z^3-3z)}\,(z+1)\,dz  \cr
&=& \frac{u^{1/4}\,\ee{-\frac{2}{3}\,u^{3/2}}}{2\sqrt{\pi}} \int \ee{-(z^3/3-u z)}\,(z/\sqrt{u}+1)\,dz  \cr
&=& \frac{u^{1/4}\,\ee{-\frac{2}{3}\,u^{3/2}}}{2\sqrt{\pi}} \left( Ai(u)+Ai'(u)/\sqrt{u}\right) \cr
\eea
where $Ai$ is the Airy function.
i.e.
\beq
f_{+,+}(u) 
= \frac{(3u)^{1/6}\,\ee{-2u}}{2\sqrt{\pi}} \left( Ai((3u)^{2/3})+Ai'((3u)^{2/3})/(3u)^{1/3}\right).
\eeq

Similarly, we have
\beq
\xi_{-,0}(z) = \frac{1}{\sqrt{3}\,(z+1)}  = \sum_{k\geq 0}\,\frac{(-1)^k}{2\,3^{(3k-1)/2}}\, \,\frac{\Gamma(3k/2)}{k!\,\Gamma(k/2)} \,\,\zeta_{+}(z)^k
\eeq
i.e.
\beq
B_{-,0;+,k} = \frac{(-1)^{k}}{2\,3^{(3k+2)/2}}\, \,\frac{\Gamma(3k/2+3/2)}{k!\,\Gamma(k/2+1/2)}
\eeq
and
\beq
\hat B_{-,0;+,k} 
= \frac{1}{2^{3(k+1)}\,3^{3k+1}}\, \,\frac{(6k+1)!!}{(2k)!}
= \frac{1}{2^{4k+3}\,3^{3k+1}}\, \,\frac{(6k+1)!!}{k!\,\,(2k-1)!!}
\eeq
This gives
\beq
f_{+,-}(u) = - \sum_k \frac{1}{u^{k+1}}\,\,\frac{1}{2^{4k+3}\,3^{3k+1}}\, \,\frac{(6k+1)!!}{k!\,\,(2k-1)!!},
\eeq
which can also be computed by Laplace transform:
\bea
f_{+,-}(u^{3/2}/3) 
&=& \frac{(u^{3/2}/3)^{1/2}\,\ee{-\frac{2}{3}\,u^{3/2}}}{2\sqrt{\pi}} \int \ee{-u^{3/2}/3(z^3-3z)}\,\frac{1}{\sqrt{3}\,(z+1)}\,\,3(z^2-1)\,dz  \cr
&=& \frac{u^{3/4}\,\ee{-\frac{2}{3}\,u^{3/2}}}{2\sqrt{\pi}} \int \ee{-u^{3/2}/3(z^3-3z)}\,(z-1)\,dz  \cr
&=& \frac{u^{1/4}\,\ee{-\frac{2}{3}\,u^{3/2}}}{2\sqrt{\pi}} \int \ee{-(z^3/3-u z)}\,(z/\sqrt{u}-1)\,dz  \cr
&=& \frac{u^{1/4}\,\ee{-\frac{2}{3}\,u^{3/2}}}{2\sqrt{\pi}} \left( -Ai(u)+Ai'(u)/\sqrt{u}\right) \cr
\eea
where $Ai$ is the Airy function.
i.e.
\beq
f_{+,-}(u) 
= \frac{(3u)^{1/6}\,\ee{-2u}}{2\sqrt{\pi}} \left( -Ai((3u)^{2/3})+Ai'((3u)^{2/3})/(3u)^{1/3}\right).
\eeq

{\bf Kernels}

The kernel is thus closely related to the Airy kernel
\bea
&& \frac{uv}{u+v}-\check B_{+,+}(u,v)  \cr
&=& uv\frac{(9uv)^{1/6}\,\ee{-2(u+v)}}{2\pi}\,\cr
&& \quad \,\frac{Ai((3u)^{2/3})Ai((3v)^{2/3})+Ai'((3u)^{2/3})/(3u)^{1/3}Ai'((3v)^{2/3})/(3v)^{1/3}}{u+v}\cr
\eea

\medskip
$\bullet$
The times are computed by:
\bea
\ee{-g_+(u^{3/2}/3)} 
&=& \frac{(u^{3/2}/3)^{1/2}}{2\sqrt{\pi}} \int \ee{-u^{3/2}/3(z^3-3z+2)}\,4(z^3-2z)\,dz  \cr
&=& \frac{2\,u^{3/4}\,\ee{-\frac{2}{3}\,u^{3/2}}}{\sqrt{3\pi}} \int \ee{-(z^3/3-uz)}\,(z^3/u^2-2z/u)\,dz  \cr
&=& \frac{2\,u^{-5/4}\,\ee{-\frac{2}{3}\,u^{3/2}}}{\sqrt{3\pi}} \int \ee{-(z^3/3-uz)}\,(z^3-2uz)\,dz  \cr
&=& \frac{2\,u^{-5/4}\,\ee{-\frac{2}{3}\,u^{3/2}}}{\sqrt{3\pi}}\,\,\left(Ai'''(u)-2uAi'(u)\right) \cr
&=& \frac{2\,u^{-5/4}\,\ee{-\frac{2}{3}\,u^{3/2}}}{\sqrt{3\pi}}\,\,\left(Ai(u)-uAi'(u)\right) \cr
\eea
i.e.
\bea
\ee{-g_+(u)} 
&=& \frac{2\,(3u)^{-5/6}\,\ee{-2u}}{\sqrt{3\pi}}\,\,\left(Ai((3u)^{2/3})-(3u)^{2/3}\,Ai'((3u)^{2/3})\right) \cr
\eea
In other words, the spectral curve class of the Ising model, is the class.
\beq
\ee{\frac{55}{12^2} \kappa_1 + \frac{3.55}{12^3}\kappa_2 + \frac{8855}{12^5}\kappa_3+\dots}
\eeq
\beq
\ee{\hat t_{+,0}} = -\,\frac{\sqrt{3}}{2}.
\eeq

\subsubsection{Example Gromov--Witten theory of $P^1$ and the Hankel function}

In \cite{Norbury2011}, Norbury and Scott claim that the Gromov--Witten invariants of $\mathbb P^1$
\beq
{\cal W}_n^{(g)}(x_1,\dots,x_n) = \sum_\mu {\cal N}_{g,\mu}(\mathbb P^1)\,\,\prod_{i=1}^n \ee{-\mu_i\,x_i}\,\,\mu_i\,dx_i
\eeq
 are computed by the invariants $W_n^{(g)}$'s of the following spectral curve:
\bea
\spcurve=({\mathbb C}^*\setminus \mathbb R_+,x,y,B) \cr
{} \cr
\left\{\begin{array}{l}
x(z)=z+1/z \cr
y(z)=\ln z \cr
B(z_1,z_2) = \frac{dz_1\otimes dz_2}{(z_1-z_2)^2}
\end{array}\right.
\eea
That curve has two branchpoints:
\beq
a_+=1 \virg a_-=-1.
\eeq
Again, the curve has the symmetry:
\beq
x(-z)=-x(z)
\virg
dy(-z)=dy(z),
\eeq
so that all what we compute for $a_+$ can easily be transposed to $a_-$.

We have:
\beq
\xi_{a_+,0}(z) = \frac{1}{z-1} = \frac{1}{\sqrt{x-2}} -\frac{1}{2}+ \sum_k \frac{(2k-1)!!\,\,(-1)^k}{2^{3k+3}\,(k+1)!}\,\,(x-2)^{k+1/2}
\eeq
i.e.
\beq
B_{a_+,0;a_+,2k-2} 
= (-1)^{k}\,\,\frac{(2k-1)!!}{2^{3k}\,\,k!}
\qquad , \quad
\hat B_{a_+,0;a_+,k-1} = (-1)^k\,\, \frac{(2k-1)!!\,\,(2k-3)!!}{2^{4k}\,\,k!}.
\eeq
Similarly we have
\beq
\xi_{a_-,0}(z)=\frac{1}{z+1} = \frac{1}{2}-\sum_k \frac{(2k-1)!!\,\,(-1)^k}{2^{3k+2}\,\,k!}\,\,(x-2)^{k+1/2}
\eeq
i.e.
\beq
B_{a_-,0;a_+,2k} 
= (-1)^{k}\,\,\frac{(2k+1)!!}{2^{3k+2}\,\,k!}
\qquad , \quad
\hat B_{a_-,0;a_+,k} = (-1)^k\,\, \frac{(2k+1)!!\,\,(2k-1)!!}{2^{4k+3}\,\,k!}.
\eeq

Therefore we have
\beq
f_{+,+}(u) = 1-\sum_{k\geq 0} \frac{\hat B_{a_+,,0;a_+,k}}{u^{k+1}}
= 1-\sum_{k\geq 1} (-1)^k\,\, \frac{(2k-1)!!\,\,(2k-3)!!}{2^{4k}\,\,k!}\,\frac{1}{u^{k}}
\eeq
\beq
f_{+,-}(u) = -\sum_k (-1)^k\,\, \frac{(2k+1)!!\,\,(2k-1)!!}{2^{4k+3}\,\,k!}\,\frac{1}{u^{k+1}} = -2 \,u\,f_{+,+}'(u).
\eeq
Again, $f_{+,+}$ and $f_{+,-}$ can be computed by Laplace transforms as integrals and are found equal to Hankel functions:
\bea
f_{+,+}(u) 
&=& \frac{u^{1/2}\,\ee{2\,u}}{2\sqrt{\pi}} \int_0^\infty \ee{-u(z+1/z)}\,\frac{1}{(z-1)}\,\,(1-1/z^2)\,dz  \cr
&=& \frac{u^{1/2}\,\ee{2\,u}}{2\sqrt{\pi}} \int_0^\infty \ee{-u(z+1/z)}\,\,\frac{z+1}{z^2}\,dz  \cr
&=& \frac{u^{1/2}\,\ee{2\,u}}{2\sqrt{\pi}} \int_0^\infty \ee{-u(z+1/z)}\,\,(1+1/z)\,\,\frac{dz}{z}  \cr
&=& \frac{u^{1/2}\,\ee{2\,u}}{4\sqrt{\pi}} \int_0^\infty \ee{-u(z+1/z)}\,\,(2+z+1/z)\,\,\frac{dz}{z}  \cr
&=& -\,\frac{u^{1/2}\,\ee{4\,u}}{4\sqrt{\pi}}\,\frac{\d}{\d u}\, \int_0^\infty \ee{-u(z+1/z-2)}\,\,\frac{dz}{z}  \cr
&=& -\,\frac{u^{1/2}\,\ee{4\,u}}{4\sqrt{\pi}}\,\frac{\d}{\d u}\, H_0(2iu) \cr
%
&=& \frac{u^{1/2}\,\ee{2\,u}}{2\sqrt{\pi}} \pi\,(-iH_0(2iu)+H_1(2iu)) \cr
\eea
where $H_0$ is the Hankel function.

Applying the formula of appendix \ref{appBuv}, we have
\bea
\hat B_{a_+,a_+}(u,v) 
&=& \frac{uv}{u+v}\,\left(1-f_{+,+}(u)f_{+,+}(v)-f_{+,-}(u)f_{-,+}(v) 
\right)  \cr
&&\cr
&=& \frac{uv}{u+v}\,\left(1-f_{+,+}(u)f_{+,+}(v)-4\, uv\,f'_{+,-}(u)f'_{-,+}(v) 
\right).
\eea

Then, we compute the times $\hat t_{a_\pm,k}$ from
\bea
\ee{-g_+(u)} 
&=& 2\,\ee{u\,x(a_+)}\,\sqrt{u/\pi}\,\, \int_{\gamma_{a_+}} dy(z)\,\,\ee{-u\,x(z)} \cr
&=& 2\,\ee{2u}\,\sqrt{u/\pi}\,\, \int_0^\infty \frac{dz}{z}\,\,\ee{-u(z+1/z)} \cr
&=& 4\sqrt{u/\pi}\,\,\int_{-\infty}^\infty\,d\phi\,\,\ee{-4u\sinh^2\phi} \qquad\qquad {\rm change\,of\,variable}\,z=\ee{2\phi} \cr
&=& 4\sqrt{u/\pi}\,\,\int_{-\infty}^\infty\,\frac{ds}{\sqrt{1+s^2}}\,\,\ee{-4us^2}  \qquad\qquad {\rm change\,of\,variable}\,s=\sinh{\phi}\cr
&=& \frac{2}{\sqrt{\pi}}\,\,\int_{-\infty}^\infty\,\frac{ds}{\sqrt{1+\frac{s^2}{4u}}}\,\,\ee{-s^2} \qquad\qquad {\rm change\,of\,variable}\,s\to s/2\sqrt u\cr
&=& \frac{2}{\sqrt{\pi}}\,\,\sum_k\,\left(\begin{array}{c}-1/2\cr k\end{array}\right)\,(4u)^{-k}\,\,\int_{-\infty}^\infty\,ds\,s^{2k}\,\,\ee{-s^2}\cr
&=& 2\,\sum_k\,\left(\begin{array}{c}-1/2\cr k\end{array}\right)\,(4u)^{-k}\,\,\frac{(2k-1)!!}{2^k}  \cr
&=& 2\,\sum_k\,\frac{(2k-3)!!}{2^k\,k!}\,(4u)^{-k}\,\,\frac{(-1)^k\,\,(2k-1)!!}{2^k}  \cr
&=& 2\,\sum_k\,\frac{\hat B_{a_1,0;a_1,k-1}}{u^k}  \cr
&=&-2\, f_{+,+}(u)
\eea

The expansion of $g_+(u)$ is
\bea
\ee{-g_+(u)} &=& -2\, f_{+,+}(u) \cr
&=& 2\,\ee{-\frac{1}{16 u}+\frac{1}{64 u^2} - \frac{25}{3\,\,2^{10}\,u^3}+ \dots} \cr
&=& 2\,\ee{-\sum_k \hat t_k u^{-k}} \cr
\eea
where the $\hat t_k$'s satisfy the recursion:
\bea
 -4\,k\,\hat t_k  &=& k\,(k-1)\,\hat t_{k-1} + \sum_{j=2}^{k-1}\, (j-1)(k-j)\,\hat t_{j-1}\,\hat t_{k-j} \cr
 && \hat t_1 = \frac{-1}{16} \, , \,\, \hat t_2 = \frac{1}{64}\, , \,\, \hat t_3 = \frac{-25}{3\,\, 2^{10}}\, , \,\, \dots \cr
\eea

We have
\beq
\hat B_{a_+,a_+}(u,v) = -\,4\,\frac{uv}{u+v}\,\left(1 + \frac{4}{uv} g_+'(u)g_+'(v)\right)\,\,\ee{-g_+(u)}\,\ee{-g_+(v)}
\eeq

\beq
\hat B_{a_+,a_+}(\mu,1/\psi) = -4\,\frac{\mu\,\ee{-g(\mu)}}{1+\mu\,\psi}\,\left(1 + \frac{4\,g_+'(\mu)}{\mu} \sum_{k} k\,\hat t_k \psi^{k+2}\, \right)\,\,\ee{-\sum_k \hat t_k \psi^k}
\eeq


\subsection{Resolved conifold}

The spectral curve of resolved conifold, is (see \cite{BKMP}):
\bea
\spcurve=(\curve,x,y,B) \cr
{} \cr
\left\{\begin{array}{l}
x(z)=-f\ln z - \ln{\frac{1-z}{1-z/Q}} \cr
y(z)=-\ln z \cr
B(z_1,z_2) = \frac{dz_1\otimes dz_2}{(z_1-z_2)^2}
\end{array}\right.
\eea
It is customary to denote
\beq
X(z) =\ee{-x(z)}
\virg
Y(z)=\ee{-y(z)}
\eeq
so that the equation of the spectral curve is
\beq
X = Y^f\,\frac{1-Y}{1-Y/Q}.
\eeq

\subsubsection{Determination of $\xi_{\pm,0}$}

We have
\beq
x'(z) = -f\,\frac{(z-a_+)(z-a_-)}{z(z-1)(z-Q)}.
\eeq
and we find:
\beq
\xi_{a_+,0}(z) = \frac{1}{\sqrt {x''(a_+)/2}}\,\frac{1}{z-a_+} 
\eeq
\bea
&& f_{++}(u)  \cr
&=& -\,\frac{f}{Q}\,\, \frac{\sqrt{u}\,\ee{ux(a_+)}}{\sqrt{2\pi\,x''(a_+)}} \,\,\int (z-a_-)\,\,z^{uf-1}(1-z)^{u-1}(1-z/Q)^{-u-1}\,dz  \cr
&=& -\,\frac{f}{Q}\,\, \frac{\sqrt{u}\,\ee{ux(a_+)}}{\sqrt{2\pi\,x''(a_+)}} \sum_k Q^{-k}\,\, \frac{\Gamma(u+k+1)}{k!\,\Gamma(u+1)}\,\,\int (z-a_-)\,\,z^{k+uf-1}(1-z)^{u-1}\,dz  \cr
&=& -\,\frac{f}{Q}\,\, \frac{\sqrt{u}\,\ee{ux(a_+)}}{\sqrt{2\pi\,x''(a_+)}} \sum_k Q^{-k}\,\, \frac{\Gamma(u+k+1)}{k!\,\Gamma(u+1)}\,\,\left(\frac{\Gamma(k+uf)\,\Gamma(u)}{\Gamma(k+uf+u)} - a_- \frac{\Gamma(k+1+uf)\,\Gamma(u)}{\Gamma(k+1+uf+u)}\right) \cr
&=& -\,\frac{f}{Q}\,\, \frac{\ee{ux(a_+)}}{\sqrt{2\pi u\,x''(a_+)}} \sum_k Q^{-k}\,\, \frac{\Gamma(u+k+1)}{k!}\,\,\left(\frac{\Gamma(k+uf)}{\Gamma(k+uf+u)} - a_- \frac{\Gamma(k+1+uf)}{\Gamma(k+1+uf+u)}\right) \cr
&=& -\,\frac{f}{Q}\,\, \frac{\ee{ux(a_+)}}{\sqrt{2\pi u\,x''(a_+)}} \sum_k Q^{-k}\,\, \frac{\Gamma(u+k+1)}{k!}\,\,\frac{\Gamma(k+uf)}{\Gamma(k+1+uf+u)} \,\left(u+(1-a_-)(k+uf)\right) \cr
\eea

Case $f=-1$:
\bea
f_{++}(u) 
&=& \,\frac{1}{Q}\,\, \frac{(a_+(1-a_+/Q)/(1-a_+))^u(Q(Q-1))^{1/4}(\sqrt Q+\sqrt{Q-1})}{\sqrt{\pi u}} \cr
&& \sum_k Q^{-k}\,\, \frac{\Gamma(u+k+1)\Gamma(k-u)}{k!^2}\,\, \,\left(u+(1-a_-)(k-u)\right) \cr
\eea

\beq
f_{+-}(u) =-\,\frac{Q^{u+uf-1}f}{a_-}\,\, \frac{\sqrt{u}\,\ee{ux(a_-)}}{\sqrt{2\pi\,x''(a_+)}} \,\,\int (z-a_-)\,\,z^{-uf}(1-z/Q)^{u-1}(1-z)^{-u-1}\,dz\,
\eeq

The times are given by
\bea
\ee{-g(u)} 
&=& \frac{\sqrt u \,\ee{ux(a)}}{2\sqrt\pi}\,\int_0^1 \frac{z^{fu}\,(1-z)^u}{(1-z/Q)^u}\,\,\frac{dz}{z} \cr
&=& \frac{\sqrt u\,\ee{ux(a)}}{2\sqrt\pi}\,\sum_k Q^{-k}\,\frac{\Gamma(u+k)}{k!\,\Gamma(u)}\, \int_0^1 z^{fu+k}\,(1-z)^u\,\,\frac{dz}{z} \cr
&=& \frac{\sqrt u\,\ee{ux(a)}}{2\sqrt\pi}\,\sum_k Q^{-k}\,\frac{\Gamma(u+k)}{k!\,\Gamma(u)}\, \frac{\Gamma(fu+k)\Gamma(u+1)}{\Gamma((f+1)u+k+1)} \cr
&=& \frac{u\sqrt u\,\ee{ux(a)}}{2\sqrt\pi}\,\sum_k Q^{-k}\,\frac{\Gamma(u+k)\,\Gamma(fu+k)}{k!\,\Gamma((f+1)u+k+1)} \cr
\eea

This is an hypergeometric function
\beq
\ee{-g(u)} 
= \frac{\sqrt u \,\ee{ux(a)}}{2\sqrt\pi}\,\,
\frac{\Gamma(fu)\Gamma(u+1)}{\Gamma((f+1)u+1)}\,\,
{}_2F_1( u,fu ,(f+1)u+1 ;Q).
\eeq

\section{Conclusion}

In this article we have shown that the invariants of any spectral curve, have an interpretation in terms of intersection numbers in a moduli space
\beq
\modsp_{g,n}^\bpt
\eeq
of maps of "colored" Riemann surfaces of genus $g$ with $n$ marked points into a discrete set $\{a_1,\dots,a_{\bpt}\}$.
This result generalizes the idea that Kontsevich--Witten intersection numbers compute the Gromov--Witten theory of a point, here instead we have the Gromov--Witten theory of a discrete set of points.

\medskip

The computation also shows that to every spectral curve $\spcurve$ is associated a certain class $\CL(\spcurve)$ in that moduli space, which in some sense generalizes the Hodge class  defined through the Mumford formula.

Descendents correspond to insertion of $\psi$ classes, and the formula in Laplace transform, looks very similar to ELSV or Mari\~no--Vafa formula:
\beq
\left({\rm Laplace}_{k_1,\dots,k_n}\,W_n^{(g)}\right)(\mu_i) \propto \left<\CL(\spcurve)\,\,\prod_{i=1}^n \frac{\sum_r \mu_i f_{k_i,r}(\mu_i)\,f_{\sigma(p_i),r}(1/\psi_i)} {1+\mu_i\,\psi_i}\right>_{\modsp_{g,n}^\bpt}
\eeq

\bigskip

{\bf Mirror symmetry}

The definition of topological recursion and invariants $W_n^{(g)}$ starts with a spectral curve, i.e. a B-model geometry, and the moduli of the spectral curve are given by a 1--form $ydx$, and a Bergman kernel $B$.

On the other side, we have the Gromov--Witten theory of a set of points, i.e. an A--model geometry, and the coefficients (the moduli) are the coefficients of $\CL$, and the coefficients of $f_{i,j}(u)$, and these are obtained by Laplace transform of the B--model moduli:

$$
\begin{array}{|r|c|c|}
\hline
&\qquad {\rm B\,model} \qquad &\qquad  {\rm A\, model} \qquad \cr
\hline
&& \cr
{\rm moduli}\qquad \, &ydx\, , \,\, B & \hat t_{a_\sigma,k}\,,\,\,\check B_{a_\sigma,d;a_{\sigma'},d'} \cr
&& \cr
\hline
\end{array}
$$
with
$$
\left({\rm Laplace}_\sigma ydx\right)(\mu) \propto \ee{-\sum_k \hat t_{a_{\sigma},k}\,\mu^{-k}}
$$
$$
\left({\rm Laplace}_{\sigma,\sigma'} B\right)(\mu,\nu) \propto \sum_{k,l} \check B_{a_\sigma,d;a_{\sigma'},d'}\,\mu^{-k}\,\nu^{-l}
$$
In other words, the mirror map, which expresses the A--moduli in terms of the B--moduli, is simply the Laplace transform.

\bigskip

{\bf BKMP conjecture ?}

BKMP conjecture \cite{Mar1,BKMP} claims that if we choose $\spcurve=\hat\CYX$ to be the mirror curve of a toric Calabi--Yau 3fold $\CYX$, then the $W_n^{(g)}$'s are the generating function of Gromov--Witten invariants of stable maps $f:\sigma_{g,n}\to \CYX$ with their boundaries on a Lagrangian submanifold $L\subset\CYX$ (to which is associated the coordinate $x$ of $\spcurve$), i.e. the Gromov--Witten theory of
$$
\modsp(\CYX,L)_{g,n}.
$$
According to what we have just found, this would mean that the Gromov--Witten theory of $\modsp(\CYX,L)_{g,n}$ actually reduces to the  the Gromov--Witten theory of a set of points $\{a_1,\dots,a_\bpt\}$, where the $a_i$'s are the invariant points of the torus symmetry of $\CYX$.

This would not be too surprising, since toric symmetry implies localization on invariant loci, and in particular near invariant points $a_i$. This well known fact has given rise to the famous "topological vertex" theory \cite{Konts1995, Diaconescu2005}.

However, BKMP conjecture is yet to be proved (it was proved so far to all genus $g$ only for $\CYX=\mathbb C^3$), and our theorem here, doesn't give directly a proof...

\section*{Acknowledgments}
I would like to thank G. Borot, A. Brini, C. Kozcaz, M. Mari\~no, M. Mulase 
for useful and fruitful discussions on this subject.
This work  is partly supported by the 
ANR project GranMa ``Grandes Matrices Al\'{e}atoires" ANR-08-BLAN-0311-01, by the European Science Foundation through the Misgam program, by the Quebec government with the FQRNT,  and the CERN for its support and hospitality.


\setcounter{section}{0}
\appendix{}
\setcounter{section}{0}


%

%





\section{Proof of theorem \ref{mainth} \label{appproof}}





\subsection{Graphical representation}

In \cite{Eynard2004, EOFg}, it was observed that the definition of $W_n^{(g)}$ can be written diagramaticaly:

\beq
W_n^{(g)}(\spcurve; z_1,\dots,z_n) = \sum_{G\in {\cal G}_{g,n}^\bpt(z_1,\dots,z_n)}\,w(G)
\eeq
where ${\cal G}_{g,n}^\bpt(z_1,\dots,z_n)$ is the set of graphs with $n$ external legs, $g$ loops, constructed as follows:

\begin{figure}[t]
\centering
\label{figexgraph13}
\includegraphics[height=3cm]{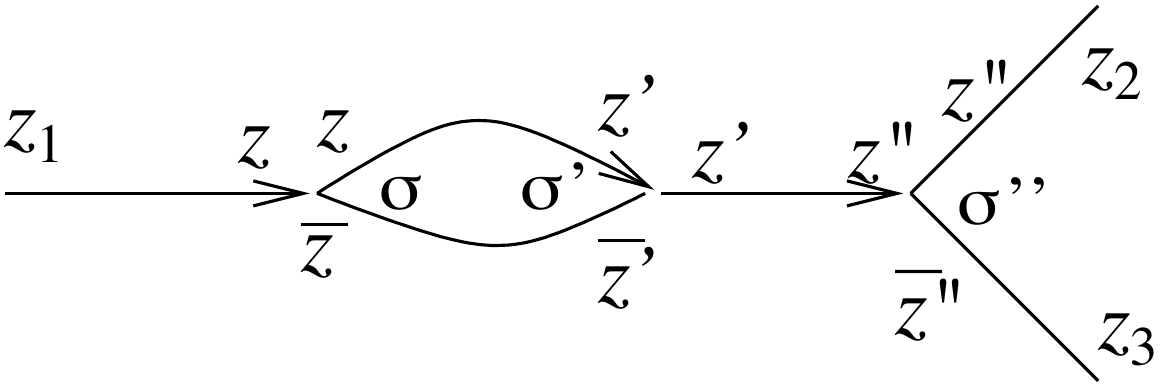} 
$$
w(G) = \Res_{z\to a_i} \Res_{z'\to a_j} \Res_{z''\to a_k}\,K(z_1,z)K(z,z')K(z',z'') B(\bar z,\bar z') B(z'',z_2) B(\bar z'',z_3)
$$
\caption{Example of a graph in $G\in {\cal G}_{1,3}^\bpt(z_1,z_2,z_3)$. $G$ has 3 external legs, 1 loop, 3 tri--valent vertices, 3 oriented edges and 3 unoriented edges.
The residue in $z''$ is computed first, then $z'$ then $z$.} 
\end{figure}

\bd
a graph $G\in {\cal G}_{g,n}^\bpt(z_1,\dots,z_n)$ if and only if:

$\bullet$ $G$ is a graph with $2g-2+n$ trivalent vertices, $n$ labeled external legs (each ending on a 1-valent vertex)

$\bullet$ $G$ has $2g-2+n$ arrowed edges forming an {\bf oriented tree}, and $n+g-1$ unoriented edges. The left and the right branch at each vertex of the tree are distinguished.

$\bullet$ each oriented edge ends on a trivalent vertex, and all trivalent vertex sit at the end of an oriented edge.

$\bullet$ each external leg but one is an unoriented edge. One external leg is at the beginning of an oriented edge, it is the root of the tree of oriented edges it has the label $z_1$. The other external legs have labels $z_2,\dots,z_n$.

$\bullet$ $G$ has $g$ internal unoriented edges, each such internal unoriented edge can connect two points only if they are on the same branch of the tree (i.e. if one is the descendent of the other).

$\bullet$ each trivalent vertex $v$ carries an index $\sigma(v)\in \{1,2,\dots,\bpt\}$.

\medskip

To associate a weight $w(G)$ to a graph $G$, we label each trivalent vertex $v$ by a spectral variable (i.e. $\in \curve$) $z_v$, and the 1-valent vertices (i.e. the root and the $n-1$ leaves of the trees are labeled by $z_1,\dots,z_n$.
This induces a labeling of edges $e=(z_{e+},z_{e-})$ of the graph in the following way:

let $v$ be a tri--valent vertex with one oriented edge $e$ with labels $(z_{e_+},z_{e_-})$ arriving on it,
and two edges (oriented or not) going out of the vertex, the left child edge $e_{\rm left}=(z_{e_{\rm left}+},z_{e_{\rm left}-})$ and the right child edge $e_{\rm right}=(z_{e_{\rm right}+},z_{e_{\rm right}-})$.
Then we have:
\beq
z_{e_-}=z_v
\virg
z_{e_{\rm left}+}=z_v
\virg
z_{e_{\rm right}+}=\bar z_v
\eeq
$$\includegraphics[height=2.5cm]{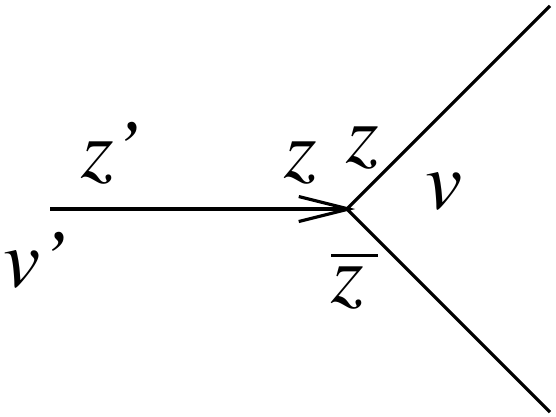} $$

- if an endpoint of an edge is a 1-valent vertex, it receives the corresponding variable $z_i$, in particular the root receives $z_1$.
\smallskip

Then the weight $w(G)$ is
\bea
w(G) =\,\,\, :\prod_{v={\rm vertices}}:\,\Res_{z_v\to a_{\sigma(v)}} \prod_{e={\rm oriented\, edges} } K(z_{e+},z_{e-})\,\, \prod_{e={\rm unoriented\, edges}}\,B(z_{e+},z_{e-})
\eea
where $:\prod:$ means that we compute the residues in a "time ordered manner", i.e. the reverse order of the arrows along the tree, i.e. "leaves first, root last".

\ed

\br
 the definition gives a special role to $z_1$, but it was shown in \cite{EOFg} that $W_n^{(g)}(\spcurve;z_1,\dots,z_n)$ is symmetric in all its variables, i.e. the result of the sum of weights is independent of which $z_i$ is chosen as the root.
\er
\medskip

{\bf Example} ${\cal G}_{0,3}^\bpt(z_1,z_2,z_3)$ contains $2\bpt$ graphs.
Each graph $G\in{\cal G}_{0,3}^\bpt(z_1,z_2,z_3)$ has 1 tri--valent vertex, 1 oriented edge and 2 unoriented edges. The oriented edge goes from the root $z_1$ to the vertex (label $z$), and the 2 unoriented edges go from the vertex to the leaves $z_2$ and $z_3$. $z_2$ is either the right or the left edge (and $z_3$ is the other). Moreover, the vertex has an index $\sigma\in\{1,2,\dots,\bpt\}$.

$$
{\cal G}_{0,3}^\bpt(z_1,z_2,z_3) = \left\{
\includegraphics[height=2.5cm]{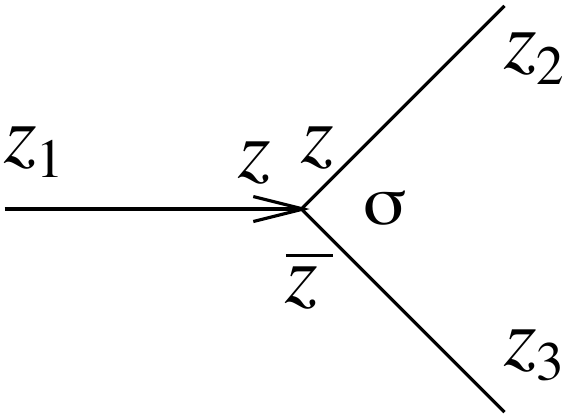} 
\virg
\includegraphics[height=2.5cm]{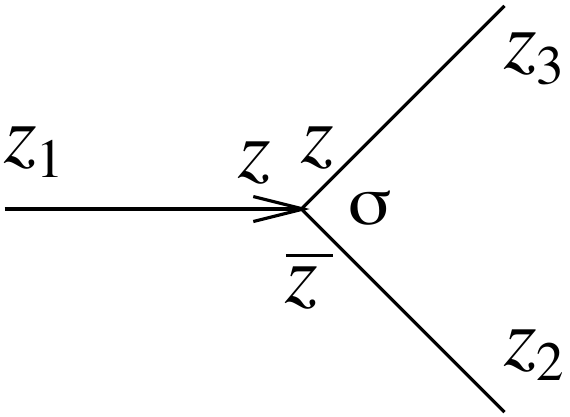} 
\right\}
$$

Eventually we have
\beq
W_3^{(0)}(\spcurve;z_1,z_2,z_3) = \sum_\sigma \Res_{z\to a_\sigma} K(z_1,z) B(z,z_2) B(\bar z,z_3)+\sum_\sigma \Res_{z\to a_\sigma} K(z_1,z) B(z,z_3) B(\bar z,z_2).
\eeq

\subsection{Structure of weights}

Since the weight $w(G)$ of a graph is evaluated by taking residues at branchpoints, it has the same structure discussed in section \ref{secstructWng}, i.e.
\beq\label{structweightgraphs}
w(G(\sigma_1,z_1;\dots,\sigma_n,z_n)) = \sum_{d_1,\dots,d_n} A(G(\sigma_1,z_1;\dots,\sigma_n,z_n);d_1,\dots,d_n)\,\,\prod_{j=1}^n d\xi_{a_{\sigma_j},d_j}(z_j)
\eeq
and the coefficients $A(G;d_1,\dots,d_n)$ is a polynomial in the $\hat B_{a_\sigma,k;a_{\sigma'},l}$'s in the $\hat t_{a_\sigma,k}$'s of total degree at most $3g-3+n$ (where $\hat B_{a_i,k;a_j,l}$ is counted with degree 1, and  $\hat t_{a_i,k}$ is counted with degree $k$), and it vanishes if $\sum_j d_j>3g-3+n$.

\subsection{Cutting graphs into clusters of constant index}

Let $G=G(i_1,z_1;\dots,i_n,z_n)\in {\cal G}_{g,n}^\bpt(z_1,\dots,z_n)$ be a graph.

Let us cut  all lines which separate vertices of different index, that gives a set of subgraphs with constant index in each subgraph:
$$
\includegraphics[height=4.5cm]{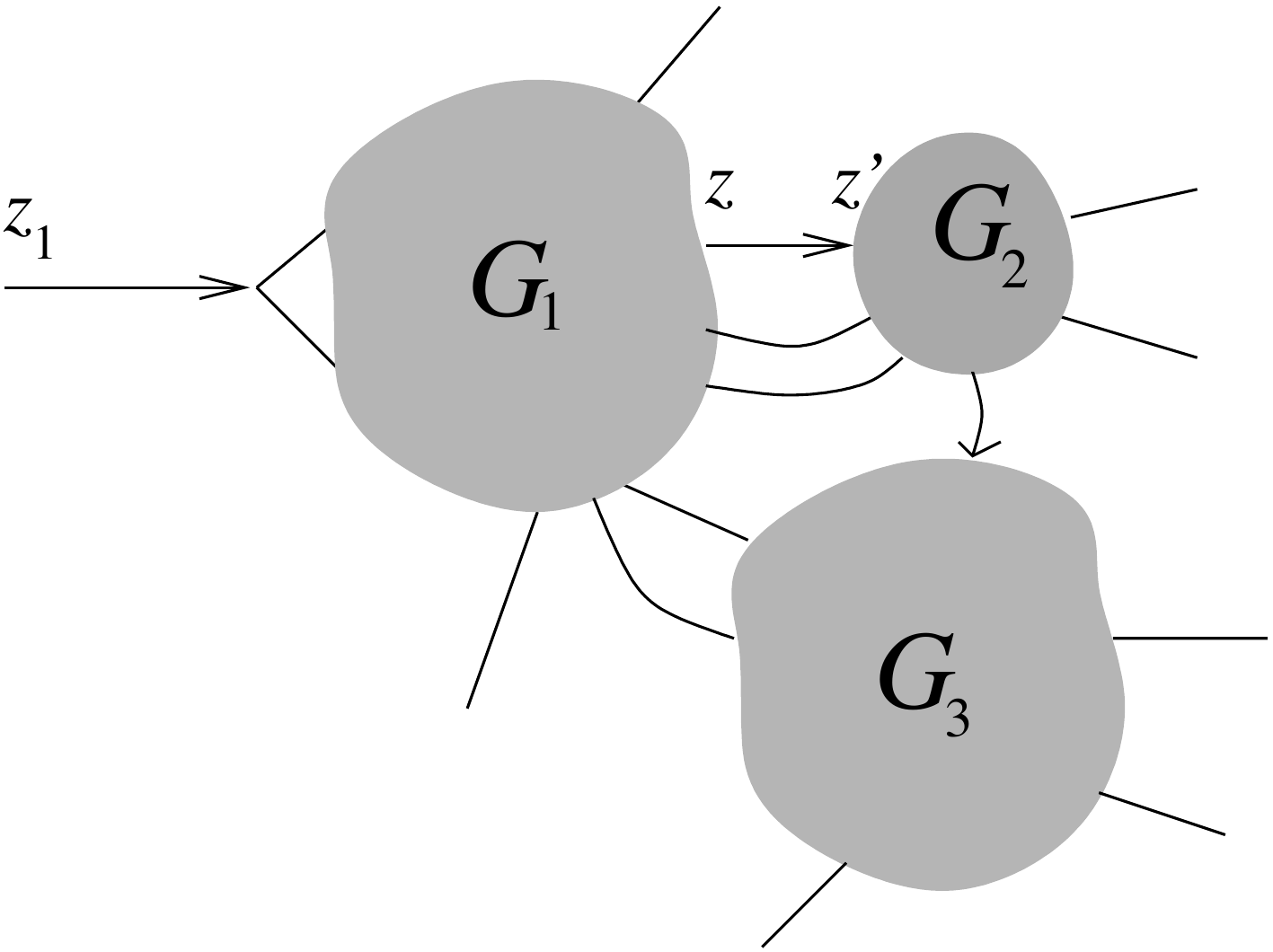} 
$$

The computation of the weight $w(G)$ involves computing residues at $z\to a_\sigma$ and $z'\to a_{\sigma'}$, i.e. taking integrals along small circles centered around  $a_\sigma$ and $a_{\sigma'}$.

If $\sigma=\sigma'$, the order of computing the two residues ($z$ or $z'$ first ?) matters, because exchanging the order means moving one circle across the other, and since $K(z,z')$ and $B(z,z')$ have poles when $z=z'$, the exchange of order produces a residue at $z=z'$ (and possibly at $z'=\bar z$ for $K(z,z')$).
But if, as we assume here, $\sigma\neq \sigma'$, the circles have different centers, and exchanging the order doesn't matter.

\medskip

So, let us compute first all residues in a subgraph $G_1$, and let $(z,z')$ be one of the external legs of graph $G_1$, i.e. it is a line separating vertices of different indexes $\sigma$ and $\sigma'$.
If $(z,z')$ is an oriented line from $z$ to $z'$, let us write
\beq
K(z,z') = \Res_{z''\to z} \Res_{z'''\to z'}\, B(z,z'') \,\,\ln{E(z'',z''')}\,\,K(z''',z')
\eeq
and If $(z,z')$ is an unoriented line between $z$ and $z'$, let us write
\beq
B(z,z') = \Res_{z''\to z} \Res_{z'''\to z'}\, B(z,z'') \,\,\ln{E(z'',z''')}\,\,B(z''',z')
\eeq
where $E(z'',z''')$ is the prime form, i.e. its second derivative with respect to $z''$ and $z'''$ is $B(z'',z''')$:
\beq\label{defprimeform}
d_{z''}\,d_{z'''}\,\ln{E(z'',z''')} = B(z'',z''').
\eeq

\smallskip
The computation of $w(G)$ involves computing a residue of the type:
\beq
\Res_{z\to a_\sigma}\,\Res_{z''\to z}\, f(z)\,B(z,z'')\,\,\ln{E(z',z''')}
\eeq
where $f(z)$ may have poles at $a_\sigma$ and possibly at other spectral variables associated to adjacent vertices in the graph $G$.
This last expression means that we should first compute the residue in $z''$ then in $z$.
This meand that $z$ is integrated around a small circle around $a_\sigma$, and $z''$ is integrated around a small circle around $z$. The circle of $z''$ can be deformed into two circles around $a_\sigma$, one exterior to the circle of $z$, and one interior:
$$
\includegraphics[height=4.5cm]{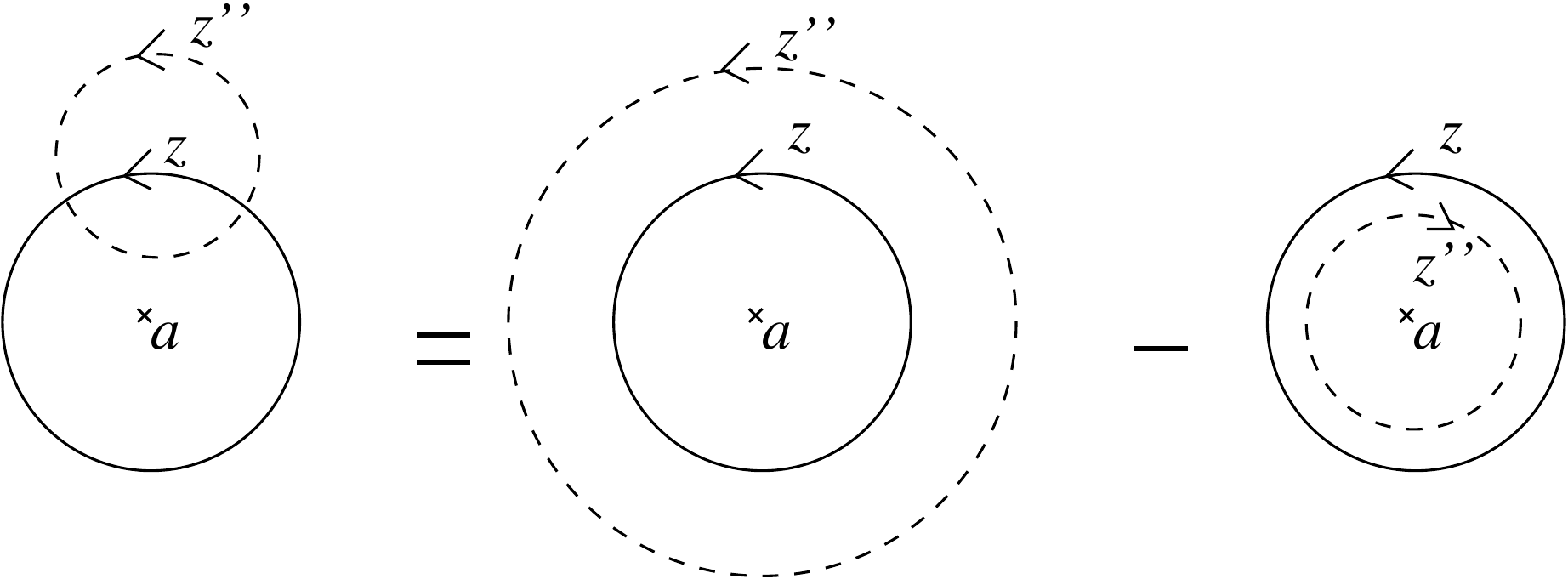} 
$$
In other words we have:
\beq
\Res_{z\to a_\sigma}\,\Res_{z''\to z}\, 
=
\Res_{z''\to a_\sigma}\,\Res_{z\to a_\sigma}\,\,
-
\Res_{z\to a_\sigma}\,\Res_{z''\to a_\sigma}\,\,.
\eeq
Since the integrand has no pole at $z''=a_\sigma$, the last residue vanishes, and therefore we may write:
\beq
\Res_{z\to a_\sigma}\,\Res_{z''\to z}\, 
=
\Res_{z''\to a_\sigma}\,\Res_{z\to a_\sigma}\,\,
\eeq
i.e. now we compute the $z$ residue first.

\smallskip
The computation of the $z$ residue, gives exactly the weight $w(G_1)$,
and since $G_1$ has all its vertices of color $\sigma$, the weight $w(G_1)$ is the same as the one computed from the spectral curve $\spcurve_{a_\sigma}$, i.e.
\beq
w(G_1) = \wrond(G_1),
\eeq
and it takes the form \eqref{structweightgraphs}
\beq\label{structweightgraphsrond}
\wrond(G_1) = \sum_{d_1,\dots,d_n} \Arond(G_1;d_1,\dots,d_n)\,\,\prod_{j=1}^n d\xi_{a_{\sigma},d_j}(z_j).
\eeq
and this has to be done for all subgraphs $G_1, G_2, \dots$ of $G$.

\smallskip
Now it remains to perform the integral over variables $z'',z'''$ associated to lines with non--constant indices in $G$, and  which now correspond to external legs of $G_1, G_2$.

Therefore we now have to compute:
\beq
C_{a_\sigma,d;a_{\sigma'},d'} = \Res_{z''\to a_\sigma}\,\Res_{z'''\to a_{\sigma'}}\,\, d\xi_{a_\sigma,d}(z'')\,\, \ln{E(z'',z''')}\,\,d\xi_{a_{\sigma'},d'}(z''')
\eeq
In this purpose we Taylor expand $\ln{E(z'',z''')}$ in the vicinity of $a_\sigma$ and $a_{\sigma'}$ using \eqref{BTaylorexpaiaj} and \eqref{defprimeform}:
\beq
\ln{E(z'',z''')} \sim \sum_{k,l}\, \frac{B_{a_\sigma,k;a_{\sigma'},d'}}{(k+1)(l+1)}\, \zeta_{a_\sigma}(z'')^{k+1}\,\zeta_{a_{\sigma'}}(z''')^{l+1}
\eeq
and
\beq
d\xi_{a_\sigma,d}(z'') \sim -\,\frac{(2d+1)!!}{2^d}\,\,\frac{d\zeta_{a_\sigma}(z'')}{\zeta_{a_\sigma}(z'')^{2d+2}} + {\rm analytical\,near\,}a_\sigma
\eeq
we have
\beq
C_{a_\sigma,d;a_{\sigma'},d'} = \frac{(2d-1)!!}{2^d}\,\frac{(2d'-1)!!}{2^{d'}}\,B_{a_\sigma,2d;a_{\sigma'},2d'} = 2\,\hat B_{a_\sigma,d;a_{\sigma'},d'} . 
\eeq

Eventually, this shows that the weight of a graph $G$ can be obtained from the weights of its subgraphs:
\beq
A(G,d_{{\rm external\,legs\,of\,}G}) = \sum_{(d_v,d_{v'}) = {\rm cut\,edges}} \prod_{(v,v')} 2\,\hat B_{a_{\sigma(v)},d_v;a_{\sigma(v')},d_{v'}}\, \prod_{G_i={\rm subgraphs}} \Arond(G_i,d_{{\rm external\,legs\,of\,}G_i})
\eeq

Notice that since the number of internal edges of a graph is $d_{g,n}=3g-3+n$, we have
\beq
d_{g,n}(G) = \#{\rm cut edges}+\sum_i d_{g_i,n_i}(G_i)
\eeq
and therefore the powers of $2$ match:
\bea\label{eqAgdprodAGi}
&& 2^{-d_{g,n}}A(G,d_{{\rm external\,legs\,of\,}G})  \cr
&=& \sum_{(d_v,d_{v'}) = {\rm cut\,edges}} \prod_{(v,v')} \hat B_{a_{\sigma(v)},d_v;a_{\sigma(v')},d_{v'}}\, \prod_{G_i={\rm subgraphs}} 2^{-d_{g_i,n_i}}\Arond(G_i,d_{{\rm external\,legs\,of\,}G_i})
\eea

\bigskip

Now it remains to perform the sum over all graphs $G$, which is equivalent to a sum over all cut edges and for given cut edges, the sum over all subgraphs $G_i$.
Since every graphs and subgraphs contain at least one trivalent vertex, and the number of trivalent vertices is $2g_i-2+n_i$, we have $2g_i-2+n_i>0$ for each subgraph $G_i$, and thus the sum contains only stable terms.
It is clear that the set of possible (cutting edges+ stable subgraphs), is in bijection with the nodal degenerations of a surface of genus $g$ with $n$ marked points.

\smallskip
From theorem \ref{th1bp}, the sum over all possible $G_i$'s with given $g_i,n_i$ and external legs with given degrees $d_k$ is
\beq
2^{-d_{g_i,n_i}} \sum_{G_i}  \Arond(G_i,\{d_k\}) = \left< \CL_{\sigma(G_i)}\prod_k \psi_k^{d_k}\right>_{g_i,n_i}
\eeq

Eventually, substituting into \eqref{eqAgdprodAGi}, gives
\bea
&& 2^{-d_{g,n}}A(G,d_{{\rm external\,legs\,of\,}G})  \cr
&=& \sum_{ {\rm nodal\,degenerations}} \prod_{{\rm nodal\,points}(q,q')} \sum_{d_q,d_{q'}} \prod_{(q,q')} \hat B_{a_{\sigma(q)},d_q;a_{\sigma(q')},d_{q'}}\, \prod \left< \CL_{\sigma(G_i)}\prod_{k=1}^{n_i} \psi_k^{d_k}\right>_{g_i,n_i} \cr
\eea
and we recognize that the right hand side is exactly what we have defined as
\beq
\int_{\modsp_{g,n}^\bpt}\, \CL\,\,\prod_{i=1}^n\,\psi(p_i)^{d_i}
\eeq
i.e. we have proved the theorem, at least in the case where $\Brond=B$ on each $\curve_{a_i}$.

\bigskip
The derivation above, relied on exchanging the order of residues when $\sigma\neq \sigma'$ thanks to the fact that small circles around $a_{\sigma}$ and $a_{\sigma'}$ don't intersect.

The exchange of order of residues can also work when $\sigma=\sigma'$ provided that we can move one circle through the other, which is possible if the integrand has no pole when circles cross each other.
In particular, let us write:
\beq
B(z,z')-\Brond(z,z') = \Res_{z''\to z} \Res_{z'''\to z'}\, \Brond(z,z'') \,\,\ln{\frac{E(z'',z''')}{\Erond(z'',z''')}}\,\,\Brond(z''',z')
\eeq

The procedure above can be repeated, the only difference is that now we also allow to cut edges with the same color on both vertices.
The coefficients $C_{a_\sigma,d;a_{\sigma},d'}$ are then given by
\beq
C_{a_\sigma,d;a_{\sigma},d'} = \Res_{z''\to a_\sigma}\,\Res_{z'''\to a_{\sigma}}\,\, d\xi_{a_\sigma,d}(z'')\,\, \ln{\frac{E(z'',z''')}{\Erond(z'',z''')}}\,\,d\xi_{a_{\sigma'},d'}(z''')
\eeq
i.e.
\beq
C_{a_\sigma,d;a_{\sigma'},d'} = 2\,\check B_{a_\sigma,d;a_{\sigma'},d'}-2\,\delta_{\sigma,\sigma'}\,\check \Brond_{a_\sigma,d;a_{\sigma},d'} . 
\eeq
This completes the proof of theorem \ref{mainth} $\,\,\square$.

\section{Case $dx=$ meromorphic}\label{appBuv}

Let $\curve$ be a compact Riemann surface of some genus $\genus$, and let $\acycle_i,\bcycle_j$ be a symplectic basis of non--contractible cycles on $\curve$ such that the Bergman kernel is normalized on $\acycle$-cycles:
\beq
\oint_{z_2\in\acycle_i} B(z_1,z_2)=0.
\eeq
Let
\beq
dS_{z_1,z_2}(z) = \int_{z'=z_2}^{z_1} B(z,z').
\eeq

In that case we have
\beq
\xi_{a_i,0}(z)  = -\, \frac{dS_{z,o}(a_i)}{d\zeta_{a_i}(a_i)}
\eeq
Since $dx$ is meromorphic, we see that $-d\xi_{a_i,d}/dx$ is a meromorphic function on $\curve$, with a pole of degree $2d+3$ at $a_i$, and simple poles at $a_j,\, j\neq i$, and which behaves near $a_i$ as $(2d+1)!!\,d^{-2-1}\,\zeta_{a_i}(z)^{-2d-3}-\hat B_{a_i,d;a_i,0}\,\zeta_{a_i}(z)^{-1} + O(1)$, and therefore, after substracting the simple poles at $z=a_j$, we see that this quantity has all the properties of $\xi_{a_i,d+1}(z)$, and thus
\beq
\xi_{a_i,d+1}(z) = -\,\frac{d\xi_{a_i,d}(z)}{dx(z)}\, - \sum_{j=1}^\bpt \hat B_{a_i,d;a_j,0}\,\xi_{a_j,0}(z)
\eeq
Taking the Laplace transform on contour $\gamma_k$ we have
\beq
f_{a_k,a_i,d+1}(u)
= -\,uf_{a_k,a_i,d}(u) - \sum_{j=1}^\bpt \hat B_{a_i,d;a_j,0}\,f_{k,j}(u)
\eeq
where
\bea
f_{a_k,a_i,d}(u)
&=& \frac{\sqrt{u}}{2\sqrt\pi}\int_{z\in \gamma_{a_k}} \ee{-u(x(z)-x(a_k))}dx(z)\,\xi_{a_i,d}(z)  \cr
&=& \delta_{i,k}\,(-1)^d \,u^d-\,\sum_{d'} \hat B_{a_i,d;a_k,d'}\,u^{-d'-1}
\eea
\beq
f_{i,j}(u) = \frac{\sqrt{u}\,\ee{ux(a_i)}}{2\sqrt{\pi}}
\,\,\int_{z\in \gamma_{a_i}}\,
\ee{-u x(z)}\,\xi_{a_j,0}(z)\,dx(z)
\eeq

Then, defining
\beq
\check B_{a_k,a_i}(u,v)  =  \delta_{i,k}\,\frac{uv}{u+v}  -\, u\sum_{d} v^{-d}\,f_{a_k,a_i,d}(u) =    \sum_{d,l}\,\hat B_{a_i,d;a_k,l}\,v^{-d}\,u^{-l}
\eeq
we get
\beq
\check B_{a_k,a_i}(u,v) = \frac{uv}{u+v}\,\left( \delta_{i,k}- \sum_{j=1}^\bpt \,f_{k,j}(u)\,f_{i,j}(v)\right).
\eeq



\bibliography{biblio}
\bibliographystyle{plain}

\end{document}